\documentclass[aps,pra,twocolumn,showpacs,superscriptaddress,longbibliography]{revtex4-1} 
\usepackage{graphicx}  
\usepackage{dcolumn}   
\usepackage{bm}        
\usepackage{amssymb}   
\usepackage{amsmath}   
\usepackage{subfigure} 
\usepackage{hyperref} 
\hypersetup{
    colorlinks=true,
    linkcolor=blue,
    citecolor=blue,      
    urlcolor=blue,
    pdftitle={High-pressure CaF2 revisited: a new high-temperature phase and the role of phonons in the search for superionic conductivity}
    bookmarks=true,
}

\begin{document}

\title{High-pressure CaF$_2$ revisited: a new high-temperature phase
  and the role of phonons in the search for superionic conductivity}

\author{Joseph R.~Nelson} \email{jn336@cam.ac.uk}
 \affiliation{Theory of Condensed Matter Group,
  Cavendish Laboratory, J.~J.~Thomson Avenue, Cambridge CB3 0HE,
  United Kingdom} 
\author{Richard J.~Needs} \affiliation{Theory of Condensed Matter Group,
  Cavendish Laboratory, J.~J.~Thomson Avenue, Cambridge CB3 0HE,
  United Kingdom}
\author{Chris J.~Pickard} \affiliation{Department of Materials Science
  and Metallurgy, University of Cambridge, 27 Charles Babbage Road,
  Cambridge CB3 0FS, United Kingdom} \affiliation{Advanced Institute
  for Materials Research, Tohoku University, 2-1-1 Katahira, Aoba,
  Sendai, 980-8577, Japan} \vskip 0.25cm

\date{\today}

\begin{abstract}
  We recently proposed a high-pressure and high-temperature
  $P\overline{6}2m$-symmetry polymorph for CaF$_2$ on the basis of
  \textit{ab initio} random structure searching and density-functional
  theory calculations
  [\href{https://doi.org/10.1103/PhysRevB.95.054118}{Phys.~Rev.~B
    \textbf{95}, 054118 (2017)}]. We revisit this polymorph using both
  \textit{ab initio} and classical molecular dynamics simulations. The
  structure undergoes a phase transition to a superionic phase in
  which calcium ions lie on a bcc-symmetry lattice (space group
  $Im\overline{3}m$), a phase not previously discussed for the
  group-II difluorides. We demonstrate that modelling this phase
  transition is surprisingly difficult, and requires very large
  simulation cells (at least 864 atoms) in order to observe correct
  qualitative and quantitative behaviour. The prediction of superionic
  behaviour in $P\overline{6}2m$-CaF$_2$ was originally made through
  the observation of a lattice instability at the harmonic level in
  DFT calculations. Using superionic $\alpha$-CaF$_2$, CeO$_2$,
  $\beta$-PbF$_2$ and Li$_2$O as examples, we examine the potential of
  using phonons as a means to search for superionic materials, and
  propose that this offers an affordable way to do so.
\end{abstract}

\vskip 0.25cm



\maketitle

\section{\label{sec:Intro}Introduction}
Calcium difluoride (CaF$_2$) has several technological applications,
and as a result its electronic structure and properties have been
widely studied
\cite{Burnett_PRB_2001,Rubloff_PRB_1972,Shi_PRB_2005,Wu_PRB_2006,Nelson_PRB_2017,Nakamura_CI_2017,Boulfelfel_PRB_2006,Kalita_PRL_2017,Dorfman_PRB_2010}.
Under ambient conditions, CaF$_2$ adopts the cubic fluorite structure
($\alpha$-CaF$_2$) with space group $Fm\overline{3}m$. This polymorph
of CaF$_2$ has a number of interesting optical properties, such as
intrinsic birefringence \cite{Burnett_PRB_2001} and a wide direct
band-gap at $\Gamma$ of 12.1 eV
\cite{Rubloff_PRB_1972,Shi_PRB_2005}. The optical gap in
$\alpha$-CaF$_2$ is calculated to increase with pressure
\cite{Wu_PRB_2006,Nelson_PRB_2017}. $\alpha$-CaF$_2$ and doped
variations thereof show good transmittance over a wide range of
wavelengths \cite{Nakamura_CI_2017}, making it an ideal material for
optical systems. At high temperatures, $\alpha$-CaF$_2$ undergoes a
superionic phase transition at \mbox{$T_c=1430$ K} at ambient
pressure, forming $\beta$-CaF$_2$, with F$^{-}$ ions as the diffusing
species \cite{Boyer_PRL_1980}. $\alpha$-CaF$_2$ is not alone in this
regard; superionic phase transitions are ubiquitous in materials with
the fluorite structure, such as PbF$_2$, SrCl$_2$, BaF$_2$, CeO$_2$
and Li$_2$O
\cite{Hull_PRB_1998,Oberschmidt_PRB_1980,Schmalzl_PRB_2007,Buckeridge_Ceria_2013,Gupta_PRB_2012}.

\begin{figure}
\centering
   \subfigure{\includegraphics[clip,scale=0.14]{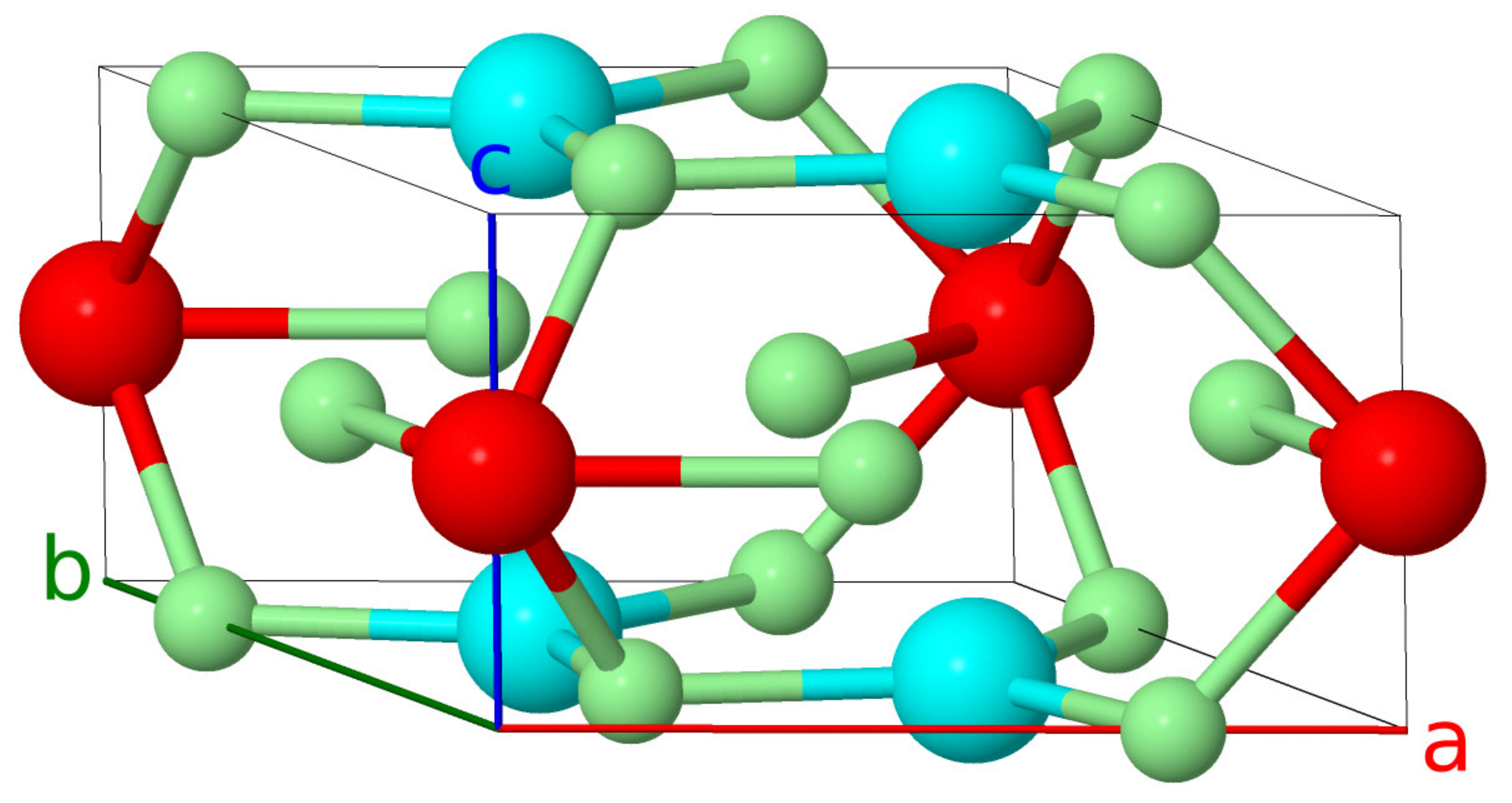}}
   \subfigure{\includegraphics[clip,scale=0.33]{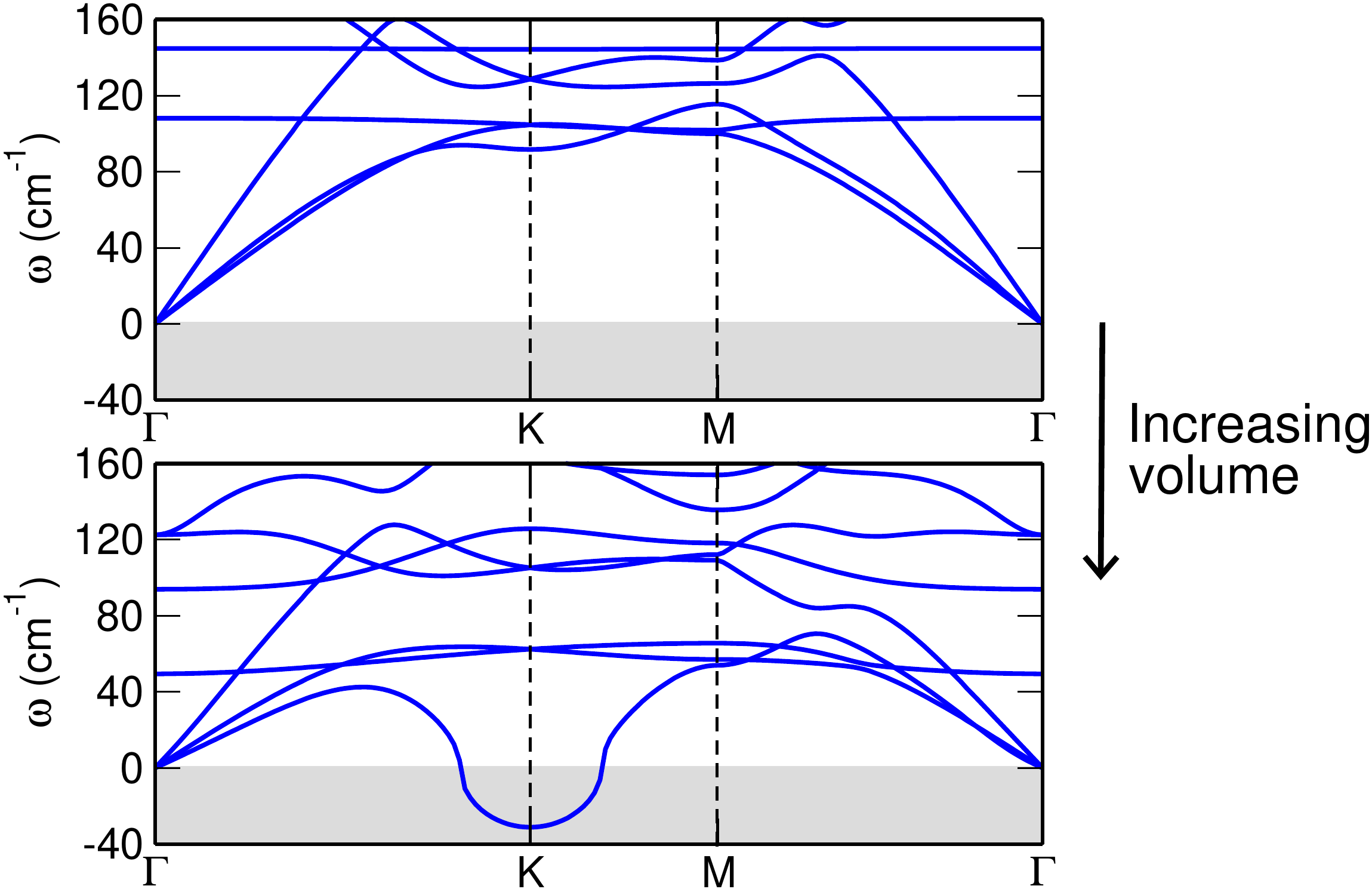}}
  \caption{\label{fig:P62m_disp} (\textit{Top}) The
    $P\overline{6}2m$-CaF$_2$ structure. Calcium atoms on the $1b$
    Wyckoff site are in red, and those on the $2c$ site are in
    blue. Fluorine atoms are shown in green. (\textit{Bottom}) Phonon
    dispersion relations in $P\overline{6}2m$-CaF$_2$ showing the
    softening of an acoustic mode at $K$ at increasing volume (see
    Ref.~\cite{Nelson_PRB_2017}).}
\end{figure}

At high pressures, $\alpha$-CaF$_2$ undergoes a phase transition to
the denser, orthorhombic cotunnite phase ($\gamma$-CaF$_2$ $-$ space
group $Pnma$) at around 9 GPa \cite{Kalita_PRL_2017}, and a further
transition to a hexagonal $P6_3/mmc$-symmetry phase at 72 GPa
\cite{Dorfman_PRB_2010}. Experimental data on high-$T$ CaF$_2$ is
scarcer. Currently available data suggests a high-$T$ modification of
$\gamma$-CaF$_2$ \cite{Mirwald_JPCS_1980,Cazorla_PRL_2014} (see also
Ref.~\cite{Mirwald_JPCS_1978}), however, these data have not yet
structurally characterized this phase. Theoretical work has proposed
that $\gamma$-CaF$_2$ melts directly at high temperature
\cite{Zhao-Yi_CPL_2008}, becomes superionic at high temperature (in
the same structure) \cite{Cazorla_JPCC_2013}, or undergoes a phase
transition to another solid phase which then becomes superionic
\cite{Cazorla_PRL_2014}. Our recent study proposed a
$P\overline{6}2m$-symmetry CaF$_2$ structure as a high-$T$
modification of $\gamma$-CaF$_2$ \cite{Nelson_PRB_2017}
(Fig.~\ref{fig:P62m_disp}). This conclusion was reached through
structure prediction calculations
\cite{Pickard_PRL_2006,AIRSS,APL_AIRSS}, treating thermodynamics
within the quasiharmonic approximation. $P\overline{6}2m$-CaF$_2$ has
the Fe$_2$P structure and is a known high-$T$ polymorph of BaCl$_2$
and BaI$_2$ \cite{BaCl2_P62m,BaI2_P62m}; this structure has also been
observed in other AB$_2$ compounds at high pressure such as TiO$_2$
\cite{Dekura_PRL_2011} and ZrO$_2$ \cite{Nishio_PCM_2015}. Whether
$P\overline{6}2m$-CaF$_2$ is a feasible candidate polymorph for high
pressure and temperature CaF$_2$ has been recently debated
\cite{Cazorla_JPCC_2018,Cazorla_Comment_2018,Nelson_Reply_2018}.

It was also proposed in Ref.~\cite{Nelson_PRB_2017} that an unstable
phonon mode could drive a superionic phase transition in
$P\overline{6}2m$-CaF$_2$ (Fig.~\ref{fig:P62m_disp}). This idea $-$
that certain lattice instabilities could trigger a superionic phase
transition $-$ was discussed by Boyer for $\alpha$-CaF$_2$ and the
$\alpha$-$\beta$ transition \cite{Boyer_PRL_1980}, and has been used
to infer superionic behaviour in a number of materials
\cite{Boyer_Ferro_1981,Buckeridge_Ceria_2013}.

In this paper, we revisit the proposed $P\overline{6}2m$-CaF$_2$
structure and explore its high-$T$ behaviour through
\textit{ab-initio} molecular dynamics (AIMD) simulations. We discuss
our methods first in Sec.~\ref{sec:Methods}, before moving on to our
results in Sec.~\ref{sec:Results}. In Sec.~\ref{sec:Discussion}, we
discuss our results, and we also examine links between phonon
frequencies and superionic conductivity. Finally, we report our
conclusions in Sec.~\ref{sec:Conclusions}.

\section{\label{sec:Methods}Methods}
AIMD simulations in this paper use the \textsc{cp2k} code \cite{CP2K}
and density-functional theory (DFT) with the PBE exchange-correlation
functional \cite{PBE1996}. Goedecker-Teter-Hutter (GTH)
pseudopotentials are used for Ca and F, which treat 10 and 7 electrons
as valence, respectively
\cite{GTH_pseudos_1,GTH_pseudos_2,GTH_pseudos_3}. These are used in
conjunction with DZVP `MOLOPT' Gaussian basis sets \cite{MOLOPT}. The
$\Gamma$-point is used for Brillouin-zone integration in all AIMD
simulations. Simulation cells containing 864 atoms are used in
all cases; the reason for this choice of cell size is discussed
further in Sec.~\ref{ssec:Prelims}. When compared against larger TZV2P
basis sets, the DZVP basis sets we use deliver energy differences
accurate to 4 meV/CaF$_2$, and the relative average absolute
difference in forces and pressures is less than 5\% for the two basis
sets. 

AIMD simulations in the canonical (\textit{NVT}) ensemble use
Nos\'{e}-Hoover thermostats with a time constant of \mbox{100 fs}. In
these simulations, all lattice parameters are fixed. AIMD simulations
in the constant-stress \textit{NPT} ensemble use the same thermostat
time constant and a barostat time constant of \mbox{2000 fs}, and
allow variation of all lattice parameters
($a$,$b$,$c$,$\alpha$,$\beta$,$\gamma$). Pressure is applied
hydrostatically to the simulation cell. A timestep of \mbox{1 fs} is
used throughout.

Classical molecular dynamics simulations use the \textsc{lammps} code
\cite{LAMMPS} alongside the same thermo- and barostat time constants
given above. Pair potentials used in these simulations are of the
Buckingham type and are taken from Refs.~\cite{Faraji_PRB_2017} for
CaF$_2$, and \cite{Oda_JNM_2007} for Li$_2$O.

Phonon frequency calculations use the \textsc{castep} plane-wave code
and density-functional perturbation theory \cite{CASTEP,DFPT}, in
conjunction with norm-conserving pseudopotentials generated by the
\textsc{castep} code's inbuilt `NCP17' pseudopotential library. Phonon
frequency calculations using pair potentials are performed with the
GULP code \cite{GULP}.

\section{\label{sec:Results}Results}
A stability field for $P\overline{6}2m$-CaF$_2$ was proposed at
temperatures above \mbox{1500-2000 K} and at pressures larger than
about 10 GPa, on the basis of calculations using the quasiharmonic
approximation \cite{Nelson_PRB_2017}. This section examines the
behaviour of $P\overline{6}2m$-CaF$_2$ at \mbox{20 GPa} and in the
temperature range \mbox{2500-3000 K}.

\subsection{\label{ssec:Prelims}Preliminaries}
The thermodynamic conditions \mbox{2500 K} and \mbox{20 GPa} lie
within the stability field suggested for $P\overline{6}2m$-CaF$_2$,
but not in the region where this structure is predicted to develop a
phonon instability according to Ref.~\cite{Nelson_PRB_2017}.

Prior to commencing our AIMD calculations, we use classical molecular
dynamics to investigate the simulation cell size needed to obtain
converged results, as suggested in
Ref.~\cite{Mulliner_PCCP_2015}. Here, and in what follows, we use an
orthorhombic setting (with $Z=6$) of the hexagonal unit cell of
$P\overline{6}2m$-CaF$_2$ (which has $Z=3$ - see
Fig.~\ref{fig:P62m_disp}). Convergence is judged by examining both
the qualitative and quantitative behaviour of the mean-squared
displacement (MSD) of Ca and F ions in $P\overline{6}2m$-CaF$_2$ as a
function of the number of atoms, $N$, in the simulation cell. The MSD
of a particular set of ions is calculated using:
\begin{align}
\label{eq:msd}
\mbox{MSD}(t) = \frac{1}{M}\sum_{i}|\textbf{r}_i(t)-\textbf{r}_i(0)-(\textbf{R}_{\mbox{\tiny CM}}(t)-\textbf{R}_{\mbox{\tiny CM}}(0))|^2,
\end{align}

\noindent where $\textbf{r}_i(t)$ is the position of ion $i$ in the set at time
$t$, $\textbf{R}_{\mbox{\tiny CM}}(t)$ is the center-of-mass of the
set of ions at time $t$, and the sum over $i$ runs over all ions in the
set, of which there are $M$ in total. Time-windowed averaging is not
performed.

\begin{figure}[htp]
\subfigure{\includegraphics[clip,width=0.8\columnwidth]{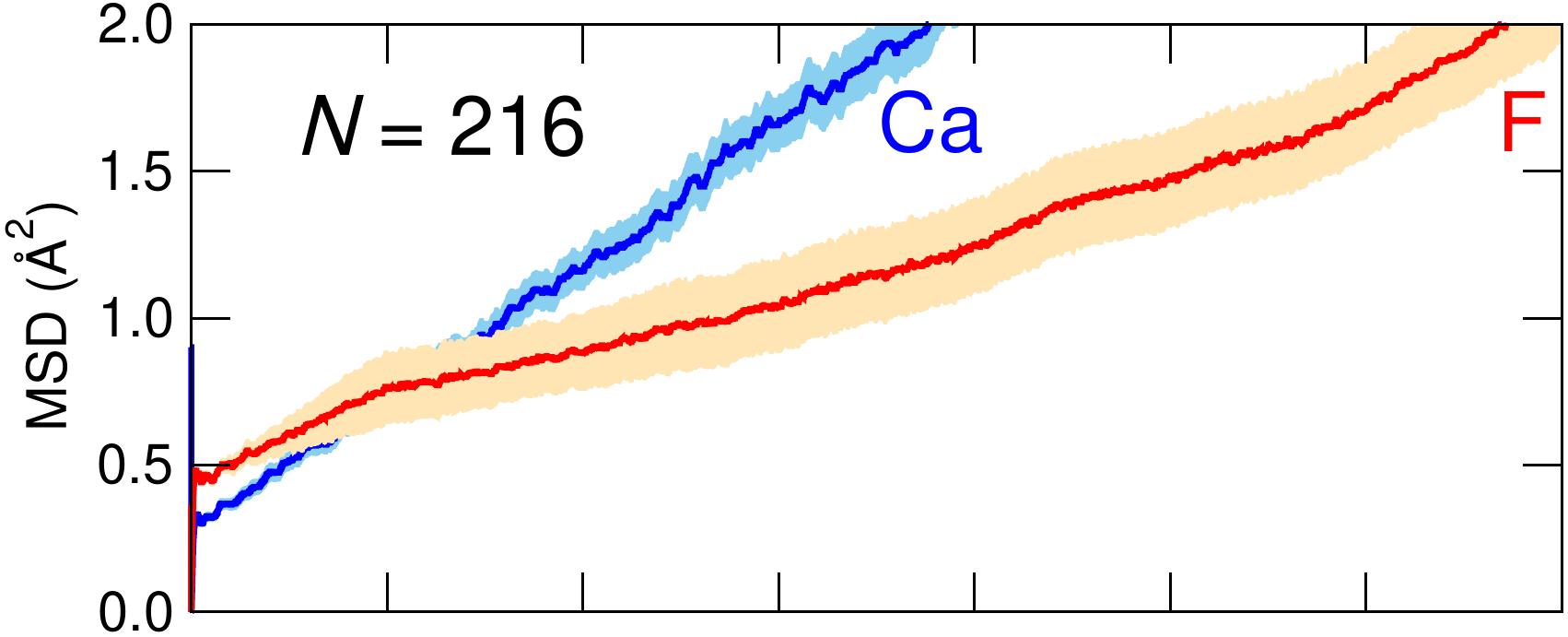}}
\subfigure{\includegraphics[clip,width=0.8\columnwidth]{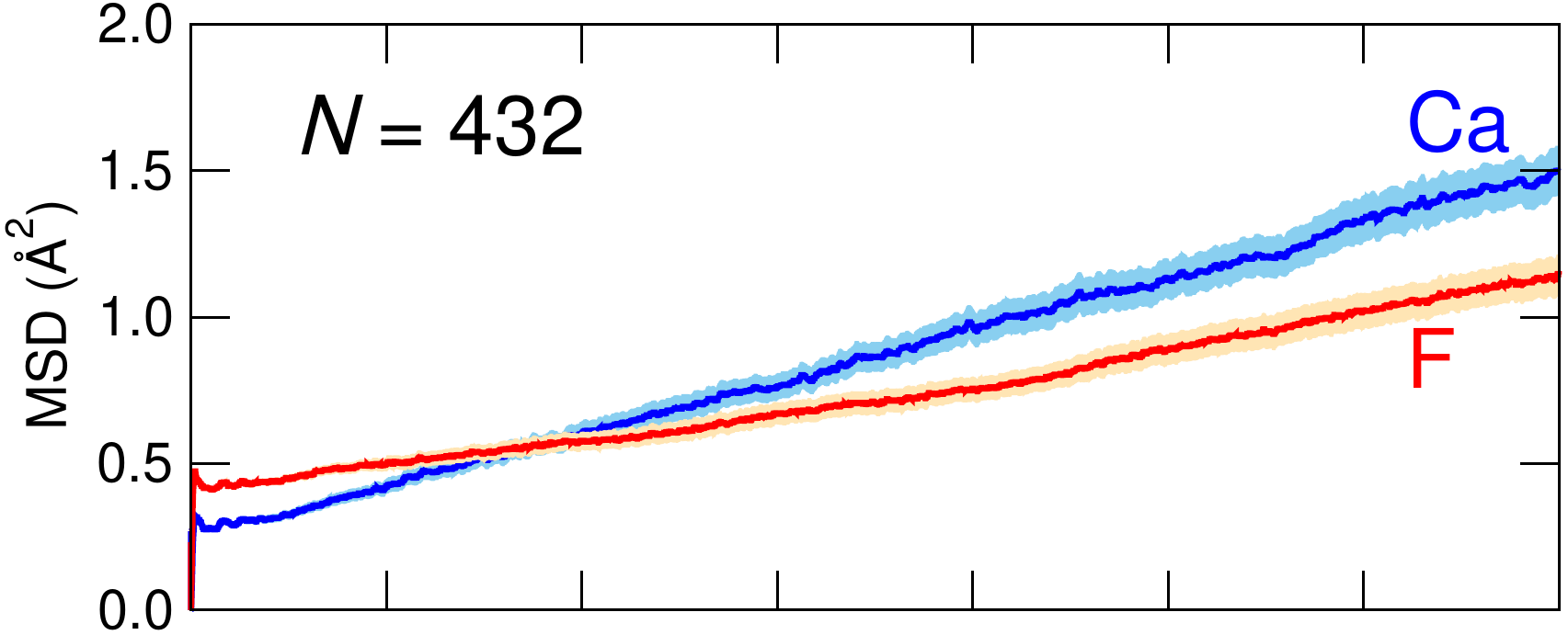}}
\subfigure{\includegraphics[clip,width=0.8\columnwidth]{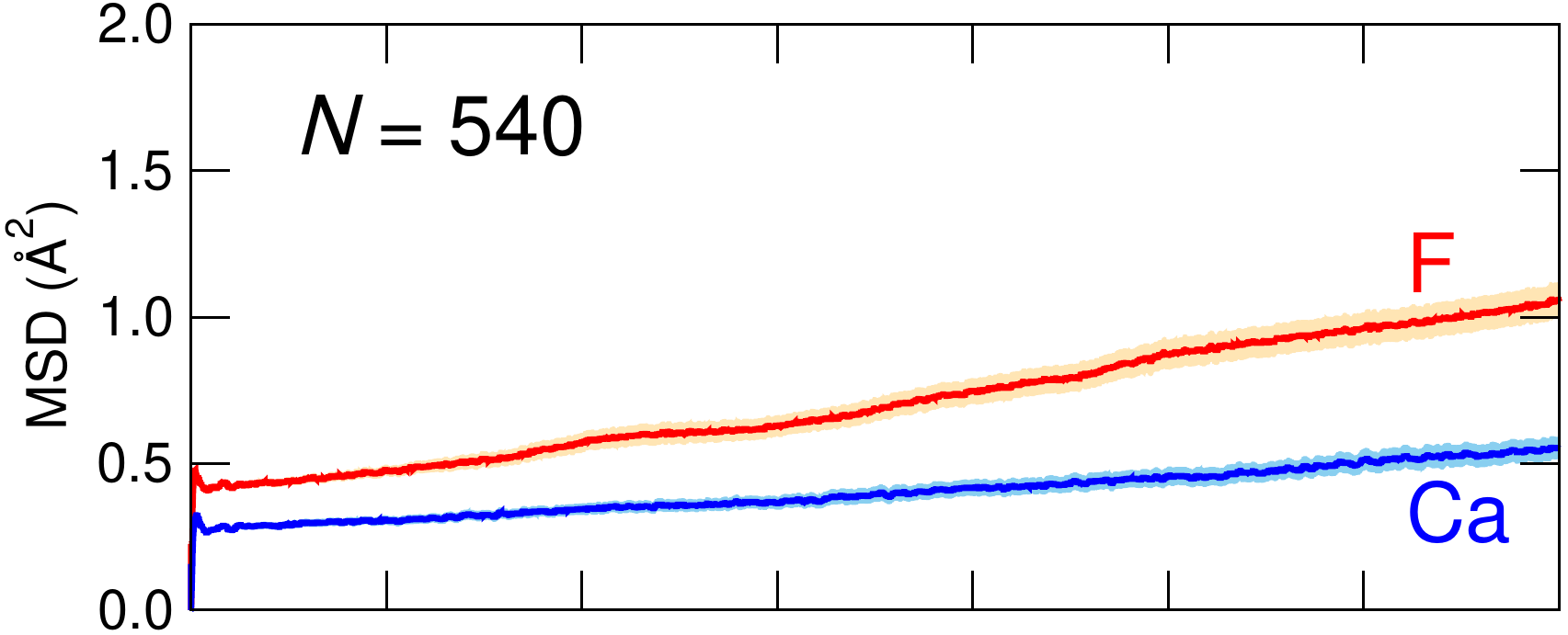}}
\subfigure{\includegraphics[clip,width=0.8\columnwidth]{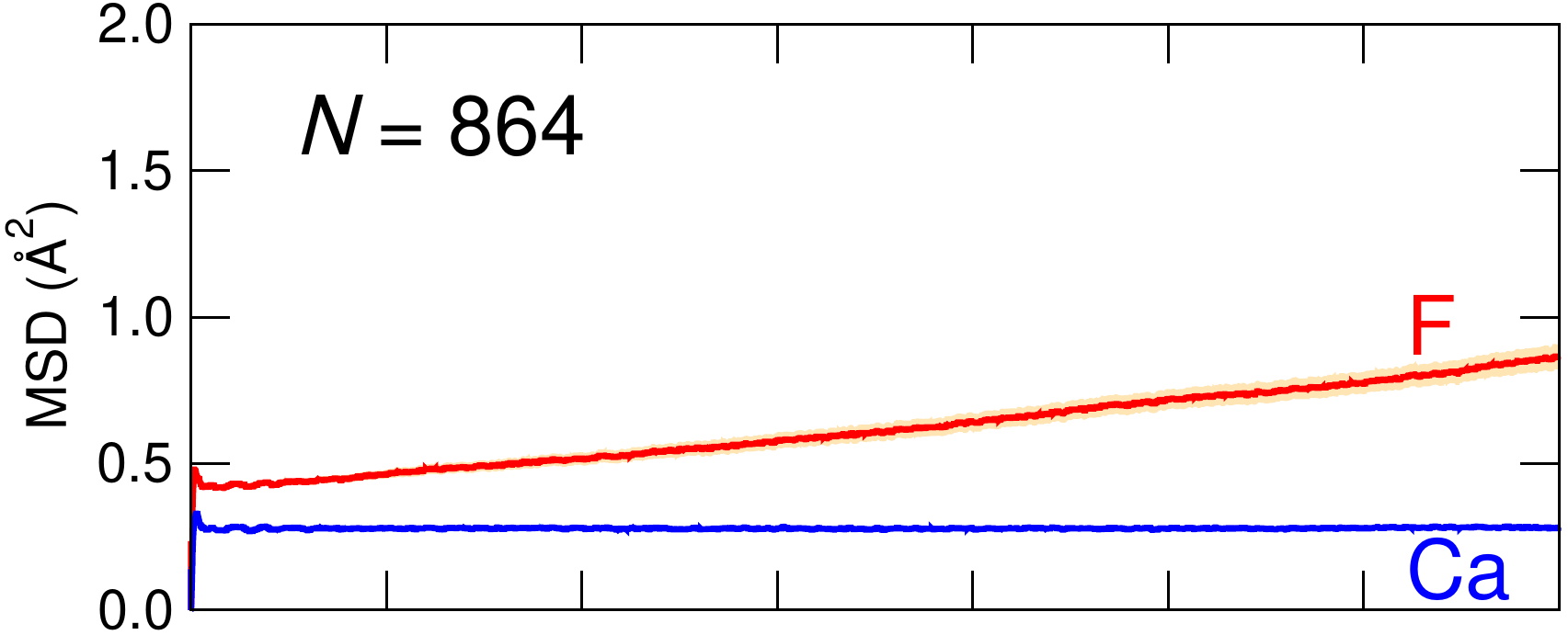}}
\subfigure{\includegraphics[clip,width=0.81\columnwidth]{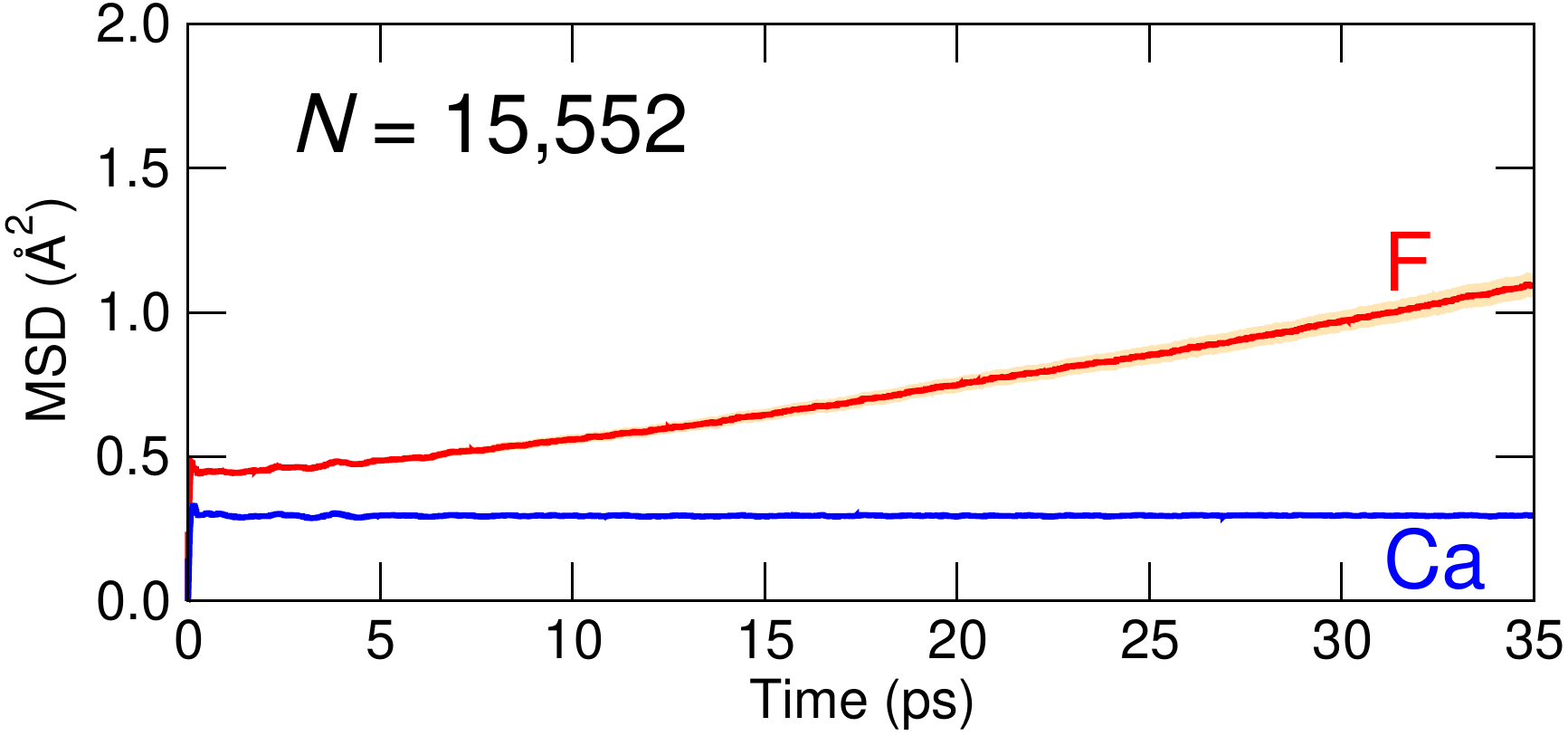}}
\caption{\label{fig:Conv}Behaviour of the mean-square displacement of
  Ca and F ions in the $P\overline{6}2m$ structure as a function of
  simulation cell size $N$ at \mbox{$T=2500$ K} and \mbox{$P=20$ GPa},
  from classical MD simulations. Light-blue and orange shaded regions
  indicate the uncertainty in the MSD for Ca and F respectively,
  obtained by averaging over multiple trajectories. Note that the
  uncertainties are small on the scale of the figures for $N=864$ and
  $N=15,552$.}
\end{figure}

Fig.~\ref{fig:Conv} shows the MSD of Ca and F ions in
$P\overline{6}2m$-CaF$_2$ at \mbox{2500 K} and \mbox{20 GPa} as a
function of cell size $N$. Supercells are constructed to be very
roughly cubic for a given $N$. Uncertainties in the MSD, as indicated
by the light-blue and orange shaded regions in Fig.~\ref{fig:Conv},
are obtained by averaging over 100 trajectories with different initial
velocities ($N$=216, 432, 540 and 864), or 20 trajectories
($N$=15,552). The results shown in Fig.~\ref{fig:Conv} were obtained in
the \textit{NVT} ensemble; cell sizes for these simulations were
obtained by first evolving a cell with $N$=15,552 in the \textit{NPT}
ensemble and then averaging the corresponding lattice parameters over
time.

The results depicted in Fig.~\ref{fig:Conv} show that the convergence
of the MSD curves with respect to simulation cell size is slow, and
surprisingly large simulation cells are required to see correct
qualitative behaviour. At small cell sizes ($N$=216), we observe
significant Ca diffusion; we have also observed such behaviour in AIMD
simulations at this cell size \cite{Nelson_Reply_2018}. There is a
significant qualitative change in the MSD curves in going from $N$=432
to $N$=540: for cells containing fewer than 432 atoms, Ca ions exhibit
a greater diffusivity than F ions (as inferred from the slope of the
MSD curves), whereas cells with more than 540 atoms show the opposite
behaviour. Referring to Fig.~\ref{fig:Conv}, we find that the MSD
curves are not qualitatively converged (as judged against $N$=15,552)
until there are at least 864 atoms in the simulation cell.

The diffusion behaviour of Ca in the $c$-direction is slowest to
converge, and the most important factor in obtaining correct
qualitative behaviour is the use of a simulation cell with a long
$c$-axis. Rather than using approximately cubic cells, we can also
obtain correct qualitative behaviour using a cell which is very
elongated in the $c$-direction but uses fewer than 864 atoms, such as
for the `$2\times 1\times 8$' cell given in the Supplemental Material
\cite{ESI}. However, when using such a cell, we find that the fluorine
diffusion coefficient is underestimated by 45\% compared to the
$N$=15,552 cell shown in Fig.~\ref{fig:Conv}. Better quantitative
agreement is obviously obtained when going to larger cells, but we
again find quite slow convergence. For our AIMD simulations in the
next section, we elect to work with the $N$=864 cell depicted in
Fig.~\ref{fig:Conv}. This size of cell shows correct qualitative
behaviour in the MSD of Ca and F, though it still underestimates the
fluorine diffusion coefficient by 33\% when compared against
$N$=15,552. The 864-atom cell is a 3$\times$2$\times$8 supercell of
the aforementioned orthorhombic $P\overline{6}2m$ unit cell. This also
means that it is commensurate with the Brillouin zone $K$-point
\cite{Monserrat_nondiag}, where a phonon instability was previously
reported at sufficiently large volumes \cite{Nelson_PRB_2017} (see
also Fig.~\ref{fig:P62m_disp}). Further MSD curves for other sizes of
simulation cell can be found in the Supplemental Material \cite{ESI}.

\begin{figure}
\centering
  \includegraphics[scale=0.43]{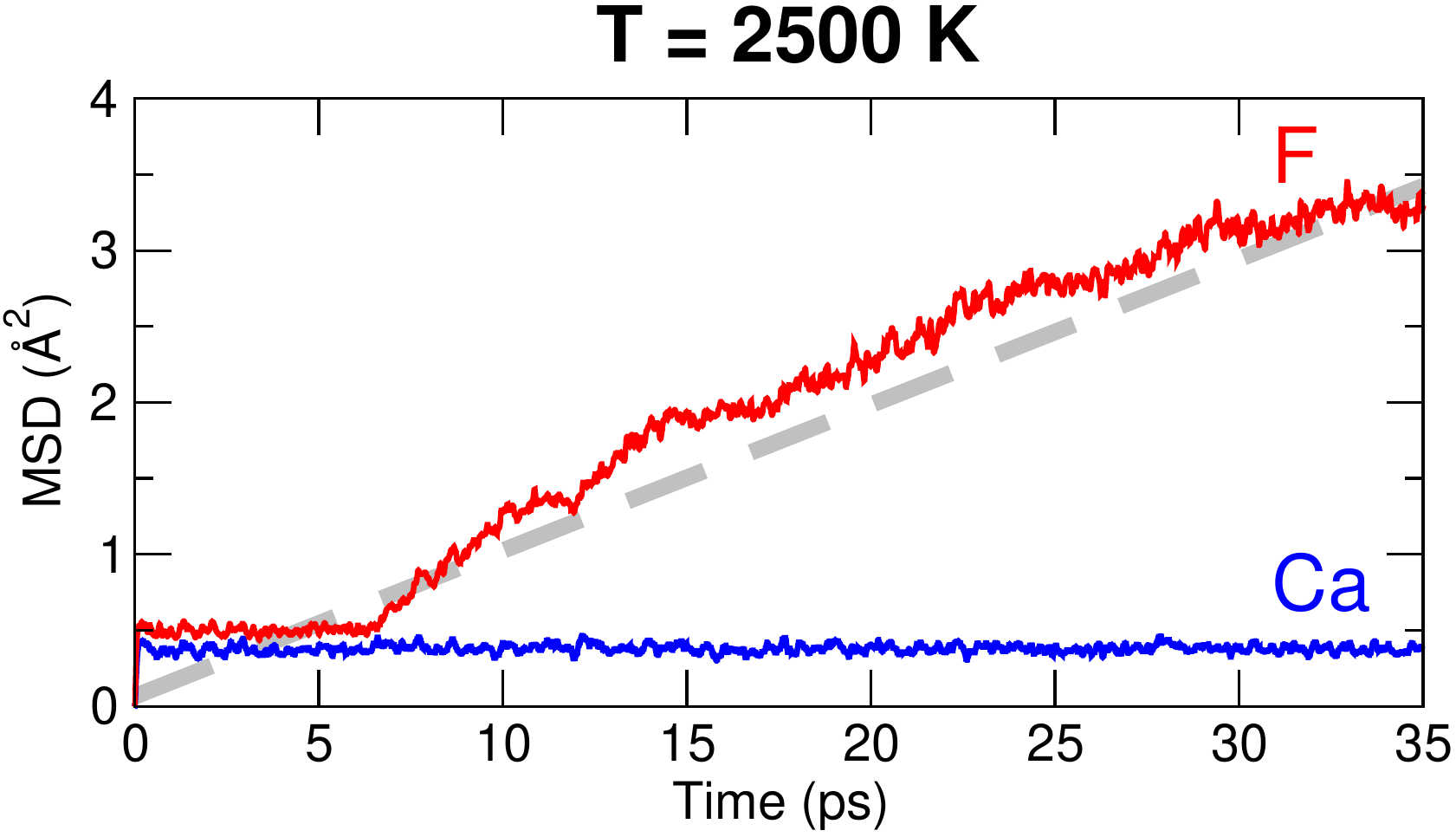}
  \caption{\label{fig:T_2500K} MSD of F and Ca ions in
    $P\overline{6}2m$-CaF$_2$ in an 864-atom AIMD-\textit{NVT}
    simulation at $T=2500$ K. The pressure is $P=19.8\pm0.4$ GPa. A
    best-fit line to the F MSD curve is shown by the thick dashed
    line.}
\end{figure}

\subsection{\label{ssec:AIMD}AIMD results}

\begin{figure*}[htp]
\centering
\begin{minipage}{.5\textwidth}
  \centering
  \includegraphics[width=.8\linewidth]{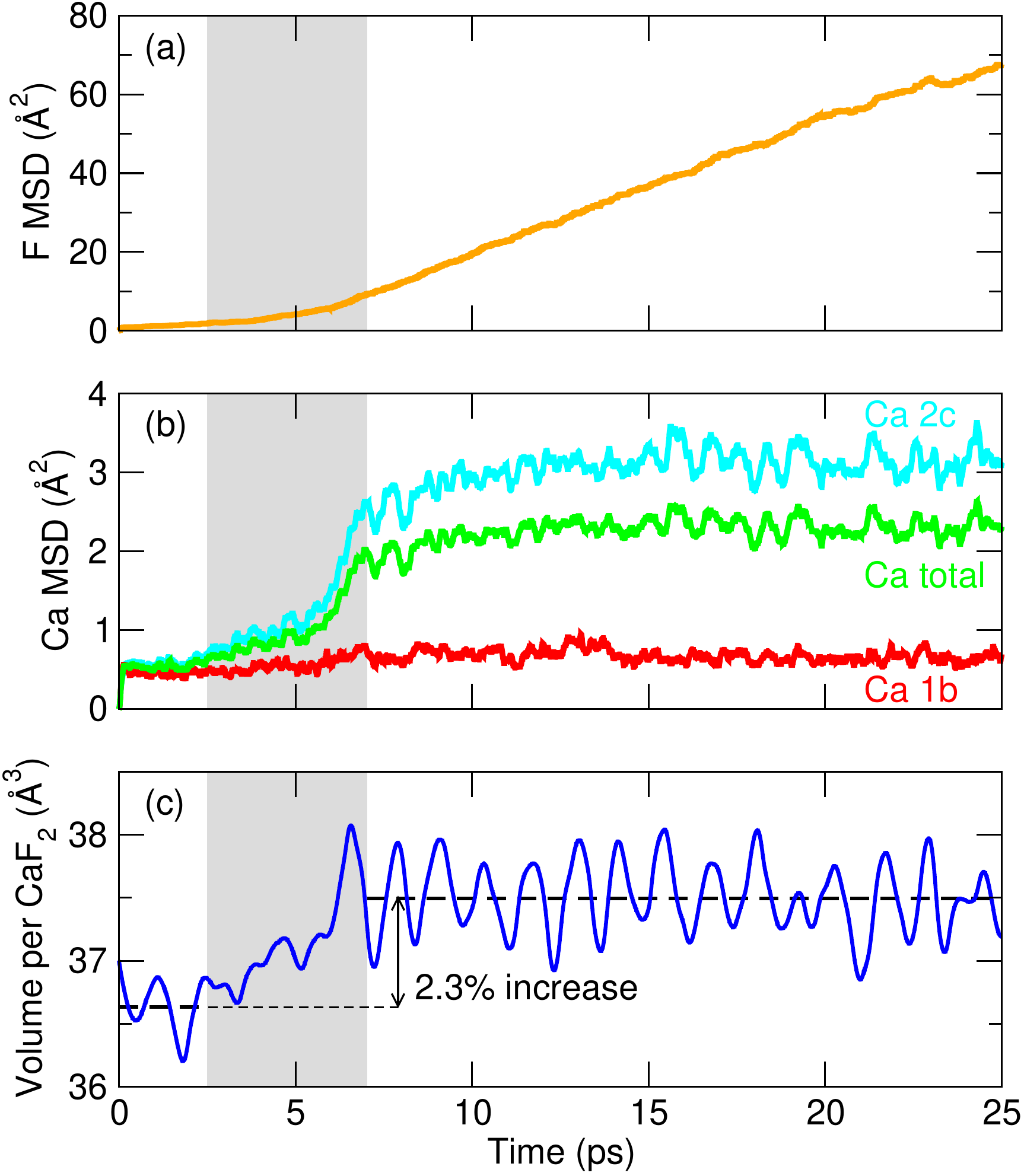}
\end{minipage}%
\begin{minipage}{.5\textwidth}
  \centering
  \includegraphics[width=.8\linewidth]{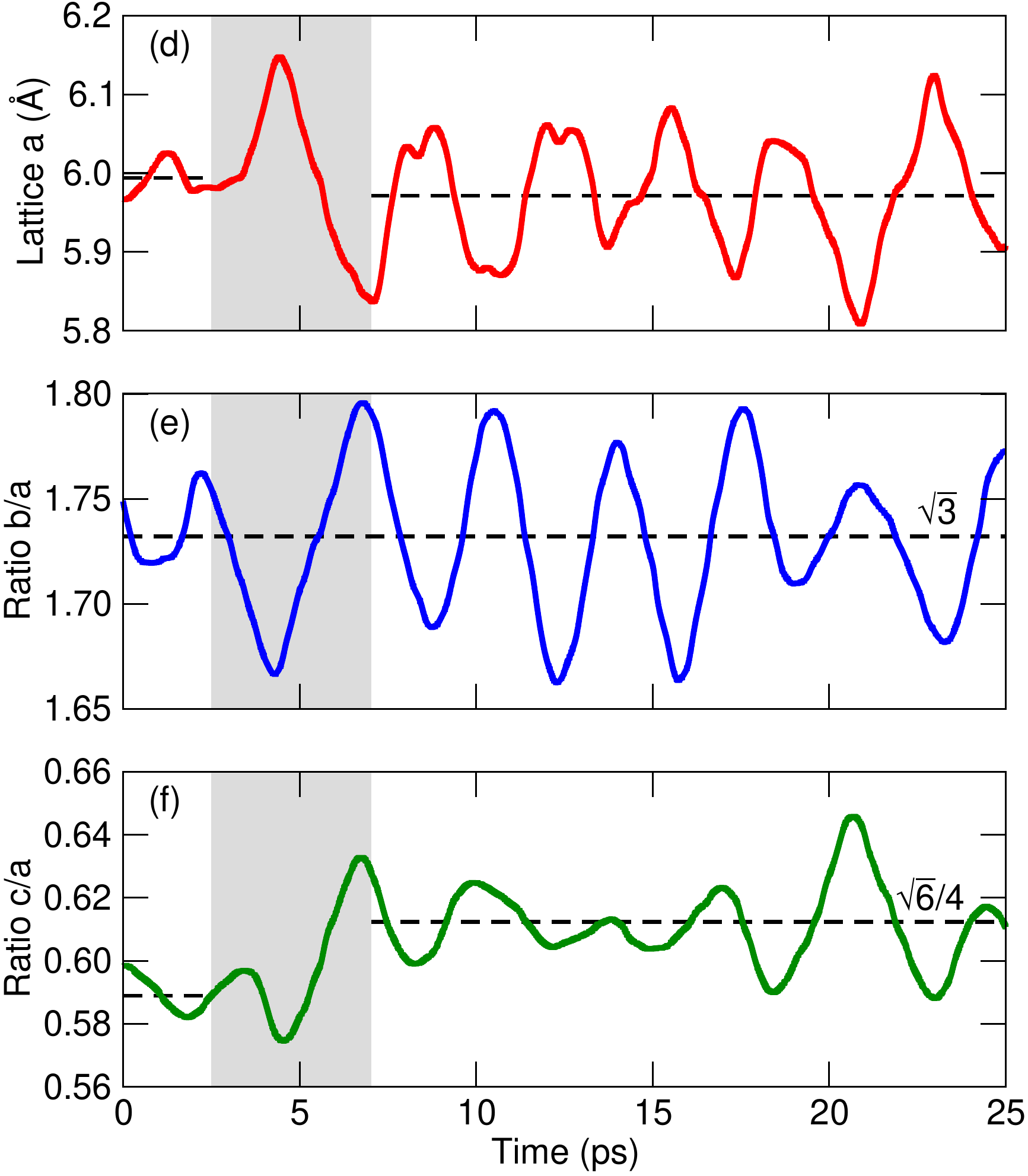}
\end{minipage}
\caption{\label{fig:T_2650K}Results of an 864-atom AIMD-\textit{NPT}
  simulation at \mbox{$T=2650$ K} and \mbox{$P=20$ GPa}. \textit{(a)}
  MSD for fluorine. \textit{(b)} MSD for calcium, split into
  contributions from calcium ions on the $1b$ and $2c$ Wyckoff
  sites. \textit{(c)} Evolution of the system volume. \textit{(d)}
  Lattice parameter $a$ for the hexagonal $P\overline{6}2m$
  cell. \textit{(e)} Ratio of the $b$ and $a$ lattice lengths in the
  orthorhombically-set cell. \textit{(f)} Ratio of the $c$ and $a$
  lattice lengths.}
\end{figure*}

\textit{T = 2500 K}. Fig.~\ref{fig:T_2500K} gives the results of an
864-atom AIMD simulation on $P\overline{6}2m$-CaF$_2$ at \mbox{2500 K}
and \mbox{20 GPa}, carried out in the \textit{NVT} ensemble. As was
the case for our classical MD simulations, the cell size was
calculated by first evolving the system in the \textit{NPT}
ensemble. The results agree qualitatively with those obtained from
classical MD simulations (Fig.~\ref{fig:Conv}) at the same cell size
($N$=864), however the diffusivity of F ions is about 8 times larger
in the AIMD simulation compared to the classical MD simulation. At
this temperature and pressure, $P\overline{6}2m$-CaF$_2$ exhibits
appreciable ionic conductivity, with F ions as the diffusing
species. The diffusion coefficient for F is $1.6\times 10^{-6}$
cm$^2$s$^{-1}$, per the slope of the F MSD curve (thick dashed line in
Fig.~\ref{fig:T_2500K}). Assuming the applicability of the
Nernst-Einstein equation \cite{Annamareddy_SciRep_2017}, the
corresponding ionic conductivity is
$\sigma\sim 10^{-2}\:\Omega^{-1}\mbox{cm}^{-1}$. No diffusion of Ca
ions is observed at this temperature and pressure. Averaging the
positions of Ca ions over the period shown in Fig.~\ref{fig:T_2500K},
and analysing the symmetry of the resulting structure using the
\textsc{c2x} code \cite{c2x} shows that Ca atoms retain their original
positions in the $P\overline{6}2m$ structure. The symmetry of the Ca
sublattice alone is $P6/mmm$. We do not observe any structural phase
transitions in $P\overline{6}2m$-CaF$_2$ at 2500 K and 20 GPa, either
in the \textit{NVT} simulation shown in Fig.~\ref{fig:T_2500K}, or in
the 20 ps long \textit{NPT} trajectory used to obtain the cell
size for the simulation shown in Fig.~\ref{fig:T_2500K}.

\textit{T = 2650 K}. A set of equilibrated atom positions and
velocities are taken from the trajectory shown in
Fig.~\ref{fig:T_2500K}, and are evolved in the \textit{NPT} ensemble
at \mbox{2650 K} and \mbox{20 GPa}. Figs.~\ref{fig:T_2650K}(a)-(f)
show the MSD of F and Ca, the volume, and the lattice parameters of
the cell as a function of simulation time.

In the \textit{NPT} ensemble, the atomic positions $\textbf{r}_i(t)$
in Eq.~\eqref{eq:msd} are affected by cell dilations. These show up as
slow undulations in the calculated MSDs. To compensate for this, the
initial positions $\textbf{r}_i(0)$ are scaled using the lattice
vectors at $t$ via
$\textbf{r}_i'(0)=[\textbf{a}(t)\;\textbf{b}(t)\;\textbf{c}(t)][\textbf{a}(0)\;\textbf{b}(0)\;\textbf{c}(0)]^{-1}[r_{ix}\;r_{iy}\;r_{iz}]^{T}$,
where $\textbf{a}(t)$, $\textbf{b}(t)$ and $\textbf{c}(t)$ are the
lattice vectors at time $t$, and $\textbf{r}_i(0)=[r_{ix}\;r_{iy}\;r_{iz}]^{T}$. $\textbf{r}_i'(0)$ is then used in place
of $\textbf{r}_i(0)$ in Eq.~\eqref{eq:msd} when the MSD is
calculated. The initial centre-of-mass position
$\textbf{R}_{\mbox{\tiny CM}}(0)$ is similarly scaled. This procedure
aids in distinguishing genuine atomic motion from that due to cell
dilations.

After a short period (2.5 ps), the cell volume increases and then
re-stabilises at around the 7 ps mark. The overall volume increase is
2.3\% (Fig.~\ref{fig:T_2650K}(c)), and occurs primarily as expansion
in the $c$-direction of the cell (3.4\%) accompanied by a small
contraction of the $a$- and $b$-axes (Fig.~\ref{fig:T_2650K}(d)). The
hexagonal ratio between the $a$ and $b$ axes, $b/a=\sqrt{3}$, is
unchanged (Fig.~\ref{fig:T_2650K}(e)). The cell remains numerically
orthorhombic over the entire trajectory shown in
Fig.~\ref{fig:T_2650K}, with $\alpha=90.2\pm0.8^{\circ}$,
$\beta=90.2\pm0.8^{\circ}$ and $\gamma=90.0\pm1.1^{\circ}$.

The change in volume is indicative of a phase transition between 2.5
and 7.0 ps, and this time interval is indicated by the grey shaded
regions in Fig.~\ref{fig:T_2650K}. Post volume-expansion, there is a
significant increase in F diffusivity
(Fig.~\ref{fig:T_2650K}(a)). Calcium ions on the Wyckoff $1b$ sites in
$P\overline{6}2m$-CaF$_2$ retain their relative positions, while those
on the $2c$ sites acquire a permanent displacement away from their
initial positions (Fig.~\ref{fig:T_2650K}(b)). There is a period,
post-volume expansion, lasting from 7.0 ps to around 15.0 ps in which
the MSD curve for Ca ions on the $2c$ site shows a slow increase
before fully stabilising.

These results are suggestive of a structural rearrangement in the
calcium sublattice, which is accompanied by a large increase in
fluorine diffusion. Averaging the calcium ion positions and lattice
parameters from 15.0 ps onward in the trajectory shown in
Fig.~\ref{fig:T_2650K} and analysing the symmetry of the resulting
structure \cite{c2x}, we find that the calcium ion sublattice is bcc
(space group $Im\overline{3}m$).

To summarise, at \mbox{2650 K} and \mbox{20 GPa}, we observe a phase
transition in $P\overline{6}2m$-CaF$_2$ in which the calcium
sublattice becomes bcc and the fluorine ions display superionic
conductivity. The structural changes and sudden increase in ionic
conductivity are characteristic of a type-I (abrupt) superionic
transition. The bcc superionic state we observe here is reasonably
well known in AB$_2$ compounds: examples include the silver
chalcogenides $\beta$-Ag$_2$S and $\beta$-Ag$_2$Se
\cite{Grier_PRB_1984,Kirchoff_PRB_1996}, and such a state is predicted
for high-pressure and high-temperature H$_2$O
\cite{Wilson_PRL_2013,Hernandez_PRL_2016}; however, we are not aware
of any previous reports of such a phase in the group-II dihalides. A
bcc superionic state has been reported in (PbF$_2$)$_{1-x}$(KF)$_x$
for $x=0.333$ \cite{Hull_PbF2KF}, with fluorine diffusing, though the
cation:anion ratio in this case is 1:1.667 as opposed to 1:2 in
CaF$_2$. Finally, we remark that this transition
($P\overline{6}2m\rightarrow Im\overline{3}m$) can be observed in
classical MD simulations, using the same interaction potentials as in
Fig.~\ref{fig:Conv}.

\begin{figure}
\centering
\includegraphics[scale=0.65]{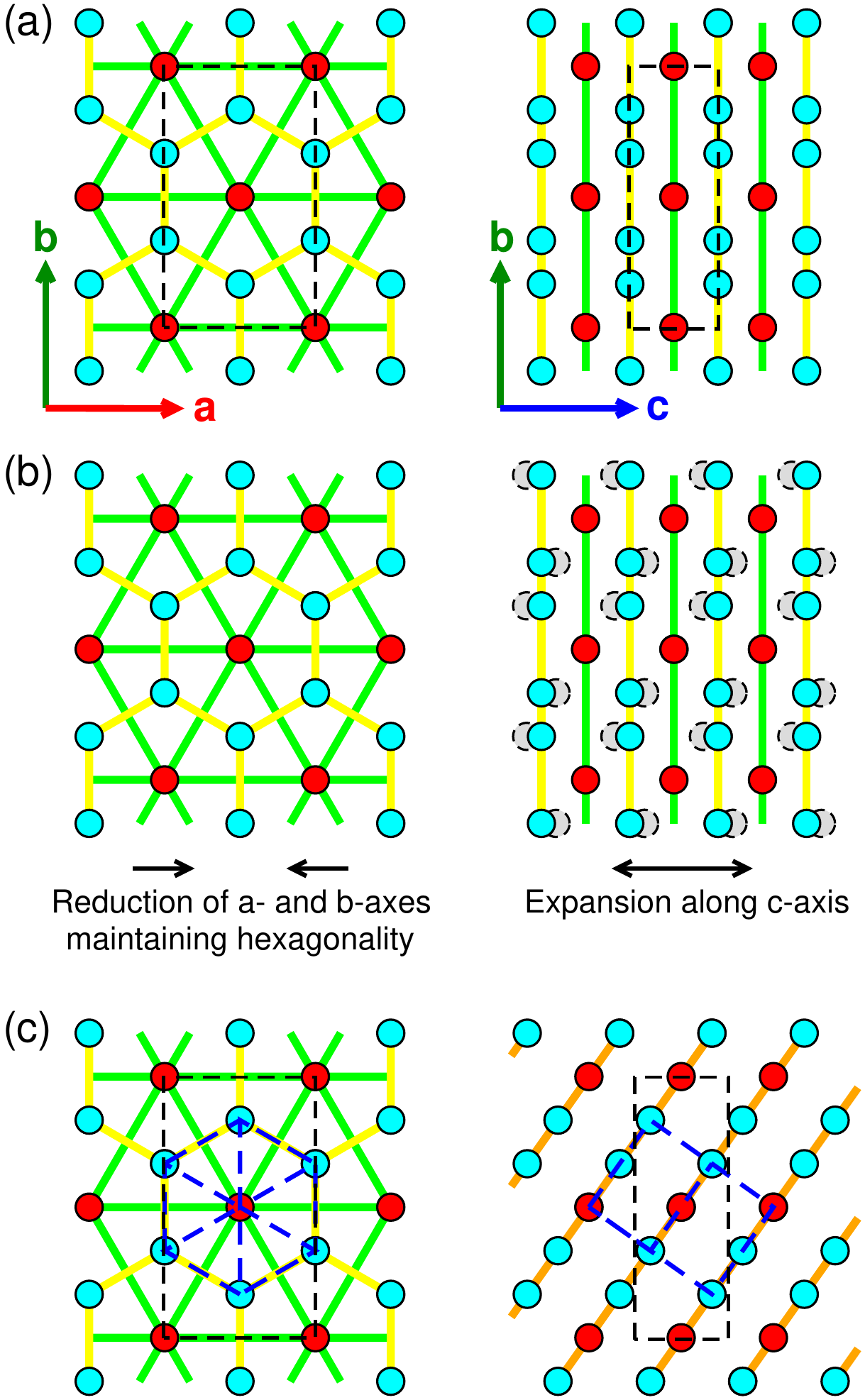}
\caption{\label{fig:P62mviews} Changes in the Ca sublattice in
  $P\overline{6}2m$-CaF$_2$. Ca ions occupying the $1b$ Wyckoff site
  are shown in red, and those occupying the $2c$ site are in
  blue. Yellow and green lines link coplanar Ca ions, in planes
  perpendicular to the $c$-axis. Fluorine ions are not
  shown. \mbox{(\textit{a}).} The pristine $P\overline{6}2m$
  structure, with black dashed lines showing the orthorhombic cell
  ($Z=6$). The hexagonal symmetry means that
  $b/a=\sqrt{3}$. \mbox{(\textit{b}).} Increasing
  temperature results in volume expansion, largely along the $c$-axis,
  with a slight reduction in the $a$- and $b$-axes. Hexagonality is
  maintained in the $a$ and $b$ axes (\textit{i.e.} $b/a=\sqrt{3}$),
  and $c/a$ increases to $\sqrt{6}/4$. Ca ions on the $2c$ positions
  (blue) are displaced away from their sites and onto the sites shown
  in grey. \mbox{(\textit{c}).}  The resulting Ca
  sublattice is bcc. The orientation of the conventional bcc unit cell
  shown by blue dashed lines.}
\end{figure}

Fig.~\ref{fig:P62mviews} shows schematically how the calcium
sublattice changes during the
$P\overline{6}2m\rightarrow Im\overline{3}m$ transition. Calcium ions
on Wyckoff $1b$ sites in $P\overline{6}2m$-CaF$_2$ (red circles in
Fig.~\ref{fig:P62mviews}) retain their relative positions, while those
on $2c$ sites (blue circles in Fig.~\ref{fig:P62mviews}) are displaced
from their initial positions. The net effect of the transition is that
these ions end up on new positions indicated by grey-dashed circles in
Fig.~\ref{fig:P62mviews}(b). Accompanying this displacement is an
expansion along the $c$-axis. Fig.~\ref{fig:P62mviews}(c) shows, using
blue-dashed lines, the bcc unit cell. To be consistent with cubic
symmetry, we would expect the orthorhombically-set $P\overline{6}2m$
cell (black dashed lines in Fig.~\ref{fig:P62mviews}(c)) to have
$b/a=\sqrt{3}$ and $c/a=\sqrt{6}/4$, which is what we observe in
Figs.~\ref{fig:T_2650K}(e) and (f).

\begin{figure}
\centering
  \includegraphics[scale=0.43]{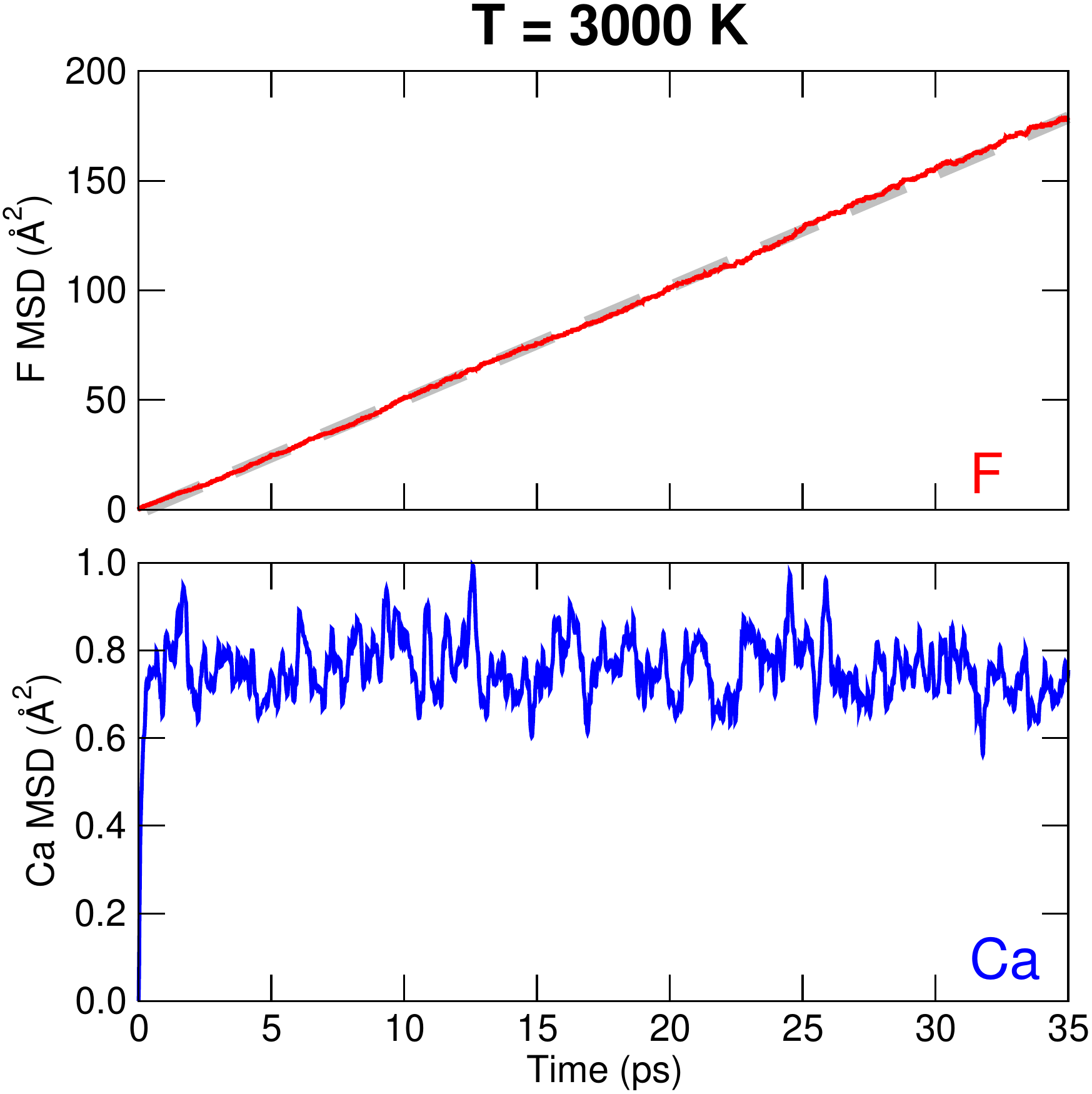}
  \caption{\label{fig:T_3000K} MSD of F and Ca ions in
    $P\overline{6}2m$-CaF$_2$ in an 864-atom AIMD-\textit{NVT}
    simulation at $T=3000$ K. The pressure is $P=19.9\pm0.4$ GPa. A
    best-fit line to the F MSD curve is shown by the thick dashed
    line.}
\end{figure}

\begin{figure}
\centering
  \includegraphics[clip,scale=0.85]{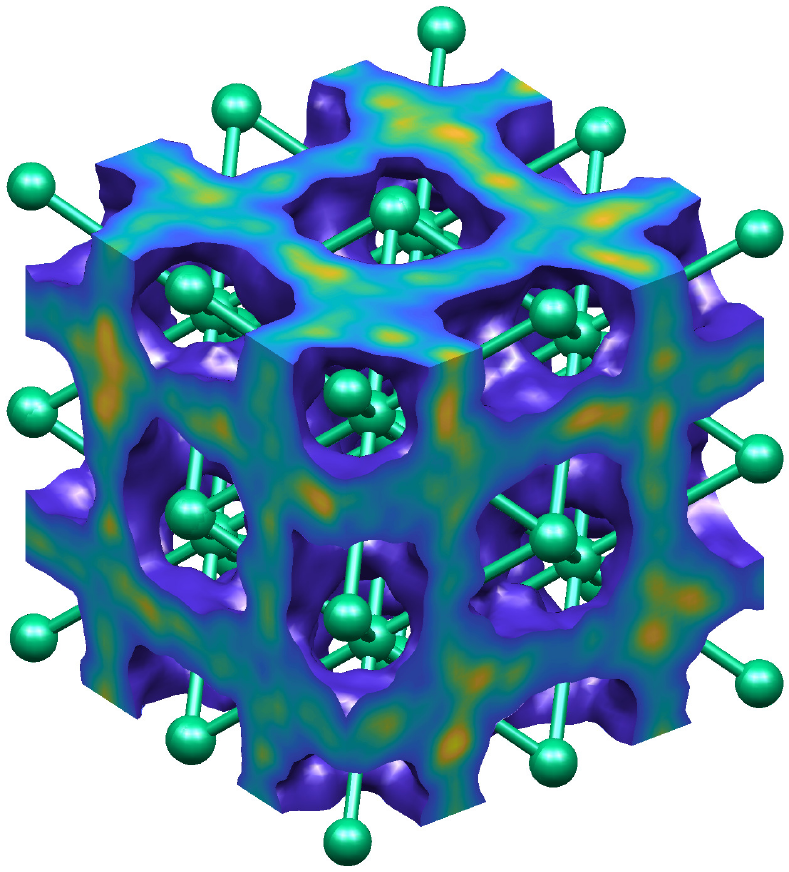}
  \caption{\label{fig:isosurfaces} Fluorine density isosurface for
    $Im\overline{3}m$-CaF$_2$ at \mbox{3000 K} and \mbox{20 GPa},
    drawn at 1$\times$ the mean fluorine density, which is 0.052
    \AA$^{-3}$. Yellow corresponds to high densities, and blue to low
    densities. Calcium ions are represented by pale blue spheres, and
    a $2\times 2\times 2$ supercell of the conventional bcc unit cell
    is shown. The conventional lattice parameter is \mbox{4.24 \AA}.}
\end{figure}

\textit{T = 3000 K}. We also carry out an AIMD simulation in the
\textit{NVT} ensemble for $Im\overline{3}m$-CaF$_2$ at \mbox{3000
  K}. The cubic lattice parameter is adjusted so that the pressure is
near 20 GPa. Fig.~\ref{fig:T_3000K} gives the MSD of calcium and
fluorine from this simulation. The calcium sublattice remains intact,
while the diffusion coefficient for F is $8.6\times 10^{-5}$
cm$^2$s$^{-1}$ (c.f.~$1.6\times 10^{-6}$ cm$^2$s$^{-1}$ at 2500 K),
corresponding to a Nernst-Einstein conductivity of
$\sigma\sim 1\:\Omega^{-1}\mbox{cm}^{-1}$
(c.f.~$\sigma\sim 10^{-2}\:\Omega^{-1}\mbox{cm}^{-1}$ at 2500
K). Fig.~\ref{fig:isosurfaces} shows the fluorine density isosurface
at this temperature and pressure, drawn at the density isovalue 0.052
\AA$^{-3}$, which corresponds to the mean fluorine density. Heatmaps
are shown in the (100), (010) and (001) planes, with yellow
corresponding to the highest density. As is fairly typical for
AB$_2$-bcc superionic conductors, we see an accumulation of density
(yellow regions in Fig.~\ref{fig:isosurfaces}) on the tetrahedral and
octahedral sites of the immobile sublattice (here calcium).

\section{\label{sec:Discussion}Discussion}

\subsection{AIMD results}
The results presented in Sec.~\ref{sec:Results} show a
high-temperature and high-pressure bcc superionic state in CaF$_2$
($Im\overline{3}m$-CaF$_2$) formed from a $P\overline{6}2m$-symmetry
polymorph at high temperature. This transition was first predicted in
Ref.~\cite{Nelson_PRB_2017} by attributing the onset of a phonon
instability at the Brillouin zone $K$ point
(Fig.~\ref{fig:P62m_disp}) to a superionic phase transition;
molecular dynamics simulations were not carried out. The AIMD
simulations in the present study indicate that the transition is both
structural and superionic, as it involves a structural rearrangement of
the calcium sublattice, as well as the onset of high fluorine
diffusivity.

We suggested previously that $P\overline{6}2m$-CaF$_2$ would be
stabilised over $\gamma$-CaF$_2$ at high temperatures, on the basis of
calculations of the $\gamma$-$P\overline{6}2m$ Gibbs free energy
difference in the QHA \cite{Nelson_PRB_2017}. Combining this with the
results of the present work, we anticipate the series of phase
transitions $\gamma$ (cotunnite)
$\rightarrow$$P\overline{6}2m$$\rightarrow$
superionic $Im\overline{3}m$ with increasing temperature, in
high-pressure CaF$_2$.  The free energy differences between
$\gamma$-CaF$_2$ and $P\overline{6}2m$-CaF$_2$ are small $-$ less than
10 meV/CaF$_2$, even at high temperature
\cite{Nelson_PRB_2017}. Examples of the
cotunnite$\rightarrow$$P\overline{6}2m$
transition in other materials, such as in ZrO$_2$
where the transition is pressure-induced \cite{Nishio_PCM_2015},
report similarly small energy differences and suggest that this
results in slow kinetics for the transition, giving rise to a
reasonably wide coexistence window for both polymorphs (cotunnite and
$P\overline{6}2m$). Such a scenario is possible in CaF$_2$.

The evidence connecting the $K$-point
phonon instability (Fig.~\ref{fig:P62m_disp}) to the superionic phase
transition seen in Fig.~\ref{fig:T_2650K} is that the soft mode
eigenvector at $K$
involves displacements of all F and Ca $2c$
ions only, and leaves stationary Ca ions on $1b$
sites \cite{Nelson_PRB_2017}, which is a feature shared by the phase
transition (Fig.~\ref{fig:P62mviews}). This is also the case for some
of the low-energy (but not soft) phonon modes appearing around
\mbox{62 cm$^{-1}$} at \textit{K} in Fig.~\ref{fig:P62m_disp}, which
show little dispersion along the $\Gamma$-$K$-$M$-$\Gamma$
path, and leave $2c$-Ca
ions stationary along significant portions of this path. In light of
this, the superionic phase transition may also involve one of these
modes, or a combination of the modes discussed here. It is perhaps
more conservative to postulate that these phonon modes drive the
observed structural phase transition in $P\overline{6}2m$-CaF$_2$
(to $Im\overline{3}m$-CaF$_2$),
but may not be involved in ion mobility, as some of the phonon modes
to be discussed in Sec.~\ref{sec:pas} are.

Under PBE exchange-correlation, the mode is not found to be completely
soft at the superionic transition volume: at 20 GPa, the transition
occurs at a volume of 1.12$V_{\mbox{\tiny static}}$, while full mode
softening is seen at 1.17$V_{\mbox{\tiny static}}$
\cite{Nelson_PRB_2017}, where $V_{\mbox{\tiny static}}$ is the
static-lattice volume of $P\overline{6}2m$-CaF$_2$ at 20 GPa. This is
not necessarily surprising, given that the
$P\overline{6}2m$-$Im\overline{3}m$ transition is first-order: if
there is a soft mode driving this transition, its frequency need not
vanish at exactly the transition temperature
\cite{VENKATARAMAN_1979,Binder}.

The AIMD results in Sec.~\ref{ssec:AIMD} differ substantially from
AIMD simulations on $P\overline{6}2m$-CaF$_2$ carried out by previous
authors \cite{Cazorla_JPCC_2018,Cazorla_Comment_2018}, where calcium
(as opposed to fluorine) diffusion was reported at \mbox{20 GPa} and
\mbox{2500 K}, and a melt state for $P\overline{6}2m$-CaF$_2$ was
reported at \mbox{20 GPa} and \mbox{3000 K}. However, it is clear from
the results given in Sec.~\ref{ssec:Prelims} that this is because the
AIMD simulations in
Refs.~\cite{Cazorla_JPCC_2018,Cazorla_Comment_2018} used simulation
cells that were not appropriately sized.

\subsection{\label{sec:pas}Phonons and superionicity in fluorite-structured ionic conductors}
\begin{figure}[htp]
\subfigure{\includegraphics[clip,width=0.8\columnwidth]{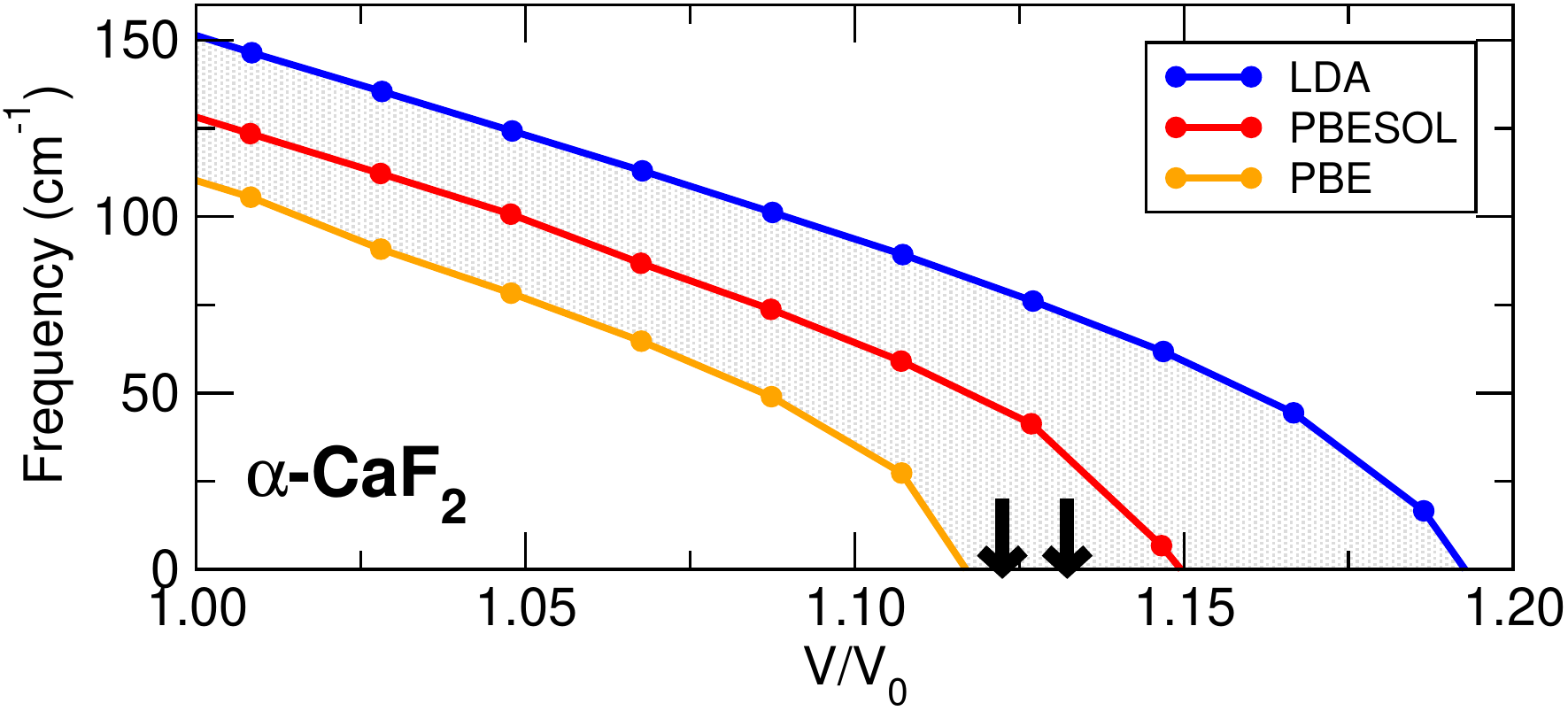}}
\subfigure{\includegraphics[clip,width=0.8\columnwidth]{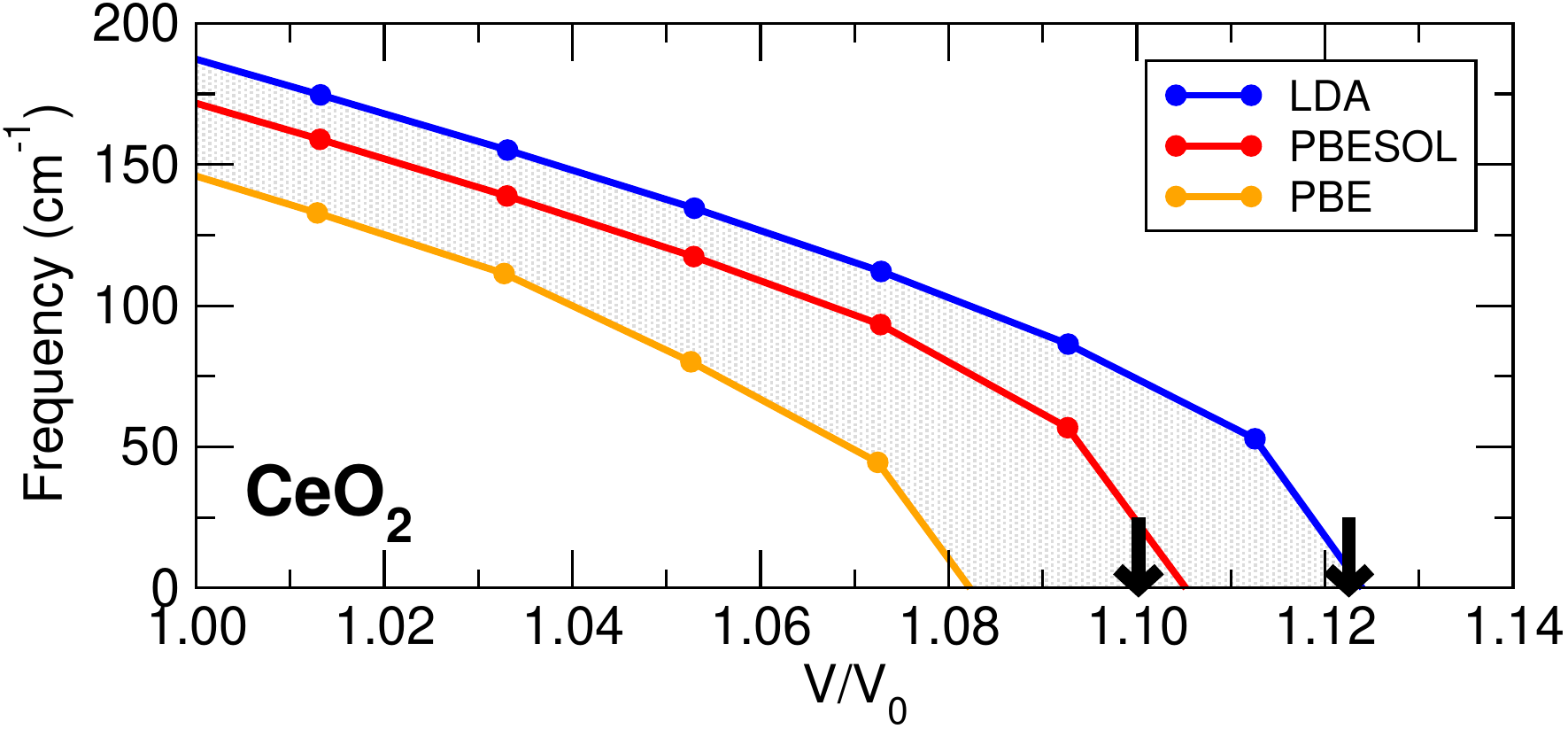}}
\subfigure{\includegraphics[clip,width=0.8\columnwidth]{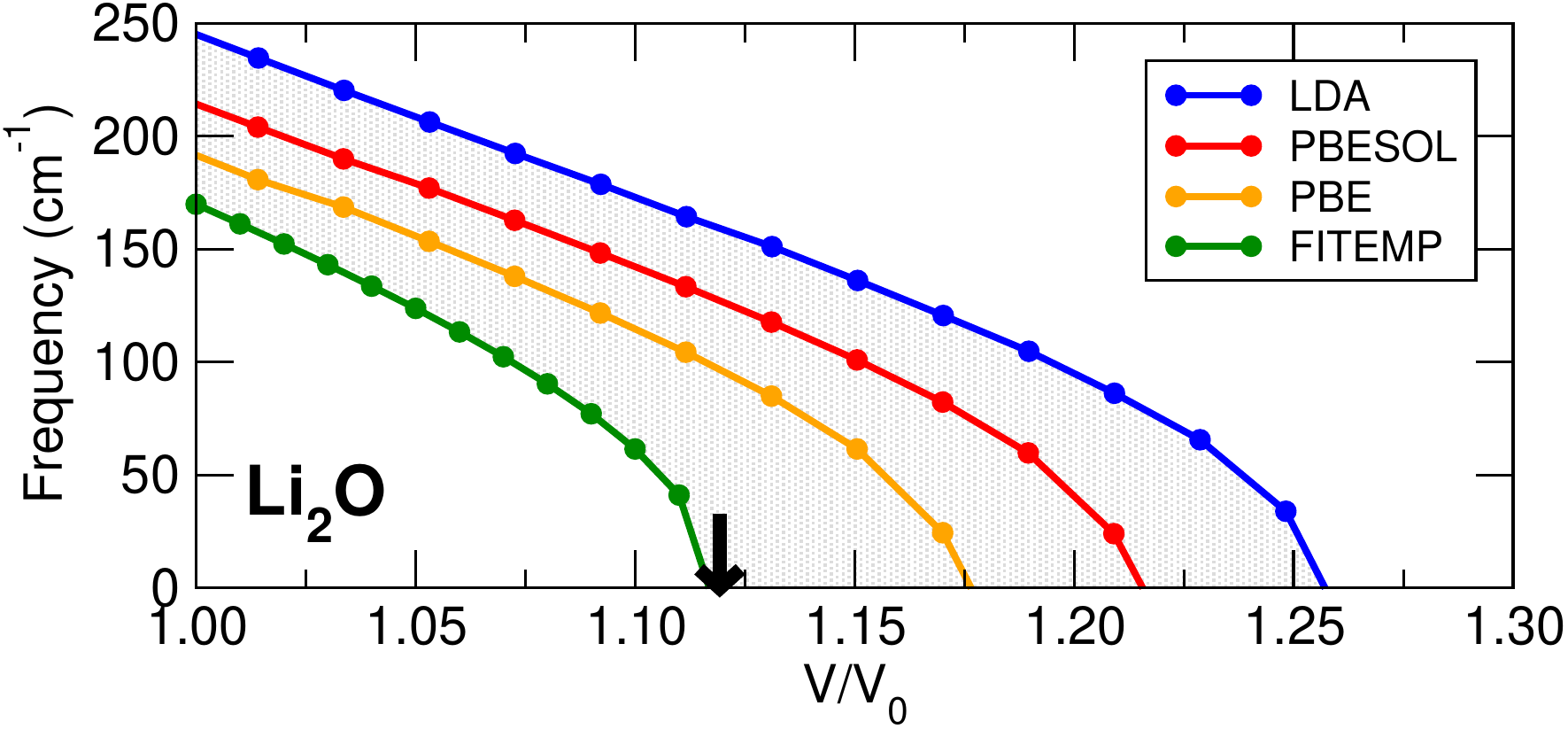}}
\subfigure{\includegraphics[clip,width=0.8\columnwidth]{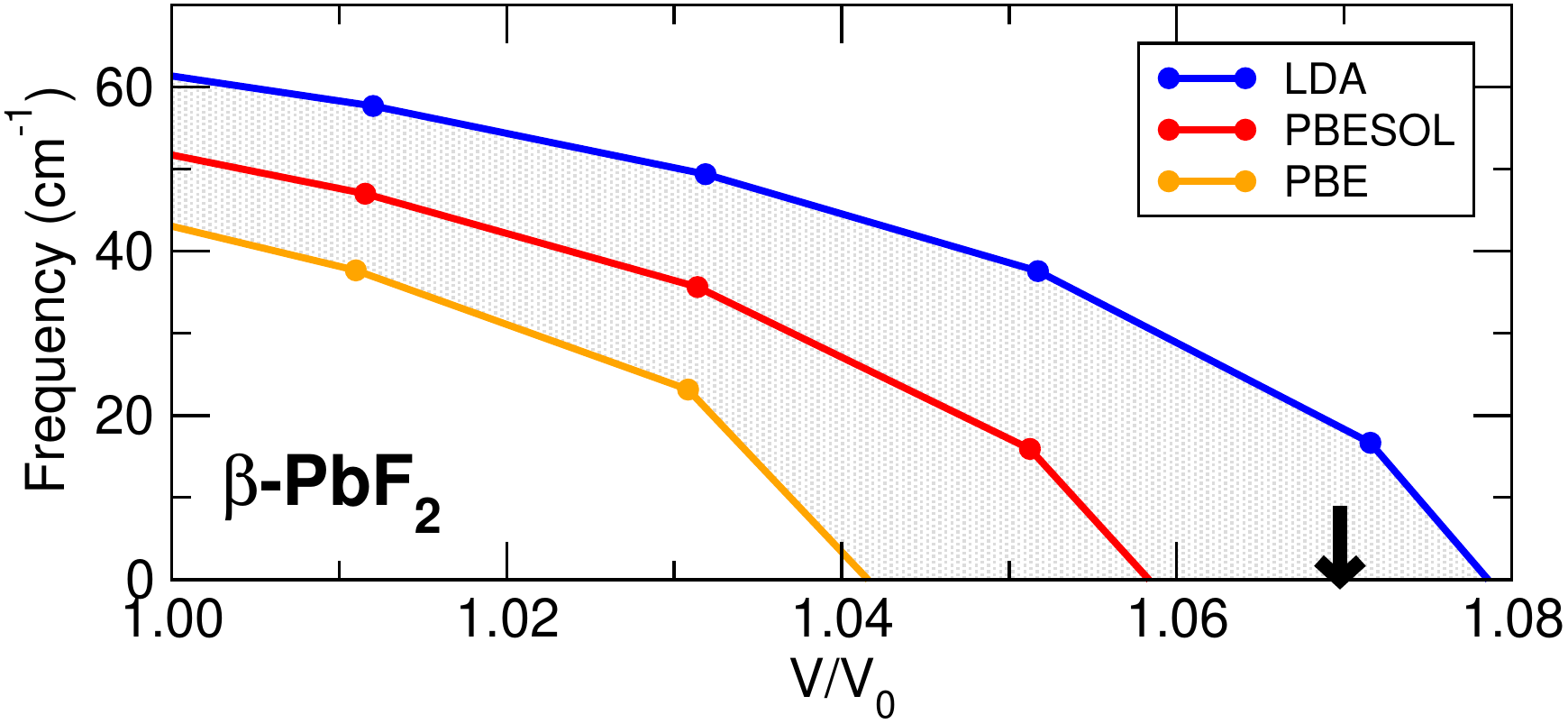}}
\caption{\label{fig:CaF2_CeO2} Softening of the $B_{1u}$ phonon mode
  in $\alpha$-CaF$_2$, CeO$_2$, and Li$_2$O, and softening of the
  $E_{u}$ phonon mode in $\beta$-PbF$_2$ with different
  exchange-correlation functionals. Frequencies $\omega$ are given as
  a function of relative volume $V/V_0$, where $V_0$ is the
  \mbox{$T=0$ K volume}. Thick vertical arrows indicate $V/V_0$
  derived from experimental results at the superionic phase
  transition; see the text for references.}
\end{figure}

As raised in Sec.~\ref{sec:Intro}, Boyer \cite{Boyer_PRL_1980}
connected a phonon instability at the Brillouin zone $X$ point in
fluorite-structured $\alpha$-CaF$_2$ to the superionic
$\alpha$-$\beta$ transition. The soft phonon mode in this case is
optical and has $B_{1u}$ mode symmetry. Buckeridge \textit{et
  al}.~\cite{Buckeridge_Ceria_2013} have, in addition to the $B_{1u}$
mode, shown a softening of the $E_{u}$ mode at $X$ in isostructural
ceria (CeO$_2$). We find it instructive to revisit a few more examples
of this phenomenon. In Fig.~\ref{fig:CaF2_CeO2}, we plot the
calculated frequencies of the $B_{1u}$ phonon mode at $X$ for
$\alpha$-CaF$_2$, CeO$_2$, and Li$_2$O, and the frequency of the
$E_{u}$ mode for PbF$_2$, for three common exchange correlation
functionals: LDA, PBE, and PBEsol \cite{LDA_1,LDA_2,PBE1996,PBESOL},
and also give results for a pair potential for Li$_2$O fitted to bulk
properties (`FIT-EMP'; Ref.~\cite{Oda_JNM_2007}). The choice of mode
($B_{1u}$ or $E_{u}$) for each compound corresponds to the mode which
first softens at increasing volume, though as $\beta$-PbF$_2$
demonstrates, both eventually soften at high enough volumes. The
reader can also refer to
Refs.~\cite{Schmalzl_PRB_2003,Buckeridge_Ceria_2013,Klarbring_Ceria_2018,Gupta_PRB_2012,Nelson_PRB_2017}
for similar calculations. These four materials are all
fluorite-structured, and all undergo type-II (continuous) superionic
transitions at sufficiently high temperatures. Frequencies are plotted
as a function of scaled volume $V/V_0$, where $V_0$ is the \mbox{$T=0$
  K} volume for each material as calculated in the quasiharmonic
approximation.

Fig.~\ref{fig:CaF2_CeO2} also shows, using thick arrows, $V/V_0$ at
the superionic phase transition as derived from experimental
data. These values are obtained as follows. For CaF$_2$, using $T_c$ =
1430 K \cite{Boyer_PRL_1980}, we deduce $V/V_0$ from the experimental
EOS data of Ref.~\cite{Schumann_thermal} (left arrow in
Fig.~\ref{fig:CaF2_CeO2}) and Ref.~\cite{Angel_EOS} (right arrow in
Fig.~\ref{fig:CaF2_CeO2}). For CeO$_2$, we assume $T_c$ = 2300 K
\cite{Kovalenko_JNM} and obtain $V/V_0$ using the experimental EOS
data of Ref.~\cite{Yashima_SSI} (left arrow) and Ref.~\cite{Omar_JACS}
(right arrow). For Li$_2$O, we take $T_c$ = 1200 K
\cite{Gupta_PRB_2012} and use the EOS data of
Ref.~\cite{Hull_JNM_1988}. Finally, for PbF$_2$, we take $T_c$ = 710 K
\cite{Hull_PRB_1998} and use the EOS data of
Ref.~\cite{Dickens_JPC_1979}. $V_0$ is either obtained directly from
experimental data, or available data on the EOS is extrapolated to
\mbox{0 K}. Data is also extrapolated to $T_c$ if the available data
does not extend to high enough temperatures.

The value of $V/V_0$ corresponding to complete mode softening varies a
fair amount between different functionals. From
Fig.~\ref{fig:CaF2_CeO2} we observe that, to within the uncertainty
introduced by the choice of exchange correlation functional (LDA, PBE
or PBEsol), there is a softening of either the $B_{1u}$ mode or $E_u$
mode coincident with the superionic phase transition in
$\alpha$-CaF$_2$, CeO$_2$, and $\beta$-PbF$_2$. Complete softening of
the $B_{1u}$ mode is found at volumes larger than the transition
volume in Li$_2$O when these three functionals are used; however, the
experimental value of $V/V_0$ at the transition agrees well with
$V/V_0$ where the $B_{1u}$ mode softens when using the aforementioned
pair potential for Li$_2$O. Functional-free techniques for calculating phonon
frequencies, such as diffusion Monte Carlo \cite{Liu_PRB_2018}, could
be used to further clarify this issue.

We close this section by re-emphasising that Fig.~\ref{fig:CaF2_CeO2},
and the discussion of phonon frequencies in this section, refer to
harmonic phonon frequencies only. Recent work on superionic CeO$_2$
\cite{Klarbring_Ceria_2018} using the temperature-dependent effective
potential method \cite{TDEP} has shown that temperature and anharmonic
effects impede a complete softening of the $B_{1u}$ phonon mode.

\subsection{Physical role of phonons}
A number of physical phenomena are coincident with the onset of a
superionic state. Examples include abrupt changes in heat capacities
\cite{Keen_PRB_2003}, an increase in the number of vacancies, number
of Frenkel or Schottky defects, or increase in occupation of
interstitial sites
\cite{Hull_PbF2KF,Hull_PRB_1998,Keen_PRL_1996,Keen_PRB_2003,Hull_JSSC_2011},
a decrease in elastic constants
\cite{Singh_PRB_1989,Gupta_PRB_2012,Buckeridge_Ceria_2013,Dickens_JPC_1979,Hull_JNM_1988},
and a softening of a particular phonon mode or modes
\cite{Boyer_SSI_1981,Boyer_PRL_1980,Buckeridge_Ceria_2013,Schmalzl_PRB_2003,Schmalzl_PRB_2007}. Samara
\cite{Samara_SSP_1984} discusses links between materials with a large
dielectric constant and superionic behaviour, and Annamareddy
\textit{et al}.~demonstrate the formation of string-like structures
comprised of conduction anions \cite{Annamareddy_SciRep_2017}. These
phenomena are not all independent, and not all of them are observed in
every superionic conductor. Of the examples given here, elastic
constants, phonon frequencies and dielectric constants can be accessed
through static calculations, as can static defect and/or vacancy
energies.

The physical role of phonons $-$ and in particular, low-energy or soft
phonon modes $-$ in superionic conduction is described in a number of
ways. Diffusing ions tend to move along directions of low curvature on
the potential energy surface (PES), and these directions should in
principle be detectable through the analysis of phonon modes, by which
one can identify low energy directions for atomic movement. Phonons
are effective in moving mobile ions toward saddle points and
contribute to diffusive jumps of mobile ions
\cite{Samara_SSP_1984,Wakamura_PRB_1997}. Energy barriers to ionic
`hopping' are expected to be smaller for `softer' or more anharmonic
lattices \cite{Samara_SSP_1984}, and low-lying or soft phonon modes
should show strong anharmonicity. As a harmonic phonon mode develops
an instability, there is a corresponding increase in amplitude of the
softening mode and a concomitant creation of a double-well energy
potential \cite{VENKATARAMAN_1979}. Such a double-well potential can
promote defect creation and lead to a higher likelihood of mobile ions
occupying interstitial sites
\cite{Buckeridge_Ceria_2013,Klarbring_Ceria_2018,Ghosh_PCCP_2016}, and
occurs in a regime in which the potential is too shallow to allow
recrystallization into another phase \cite{Boyer_Ferro_1990}. Previous
studies which have either explained or inferred superionic behaviour
on the basis of phonon modes usually proceed by analysing phonon mode
eigenvectors, and deciding whether there are soft or low-energy modes
conducive to disorder or defect creation
\cite{Buhrer_SSC_1975,Buckeridge_Ceria_2013,Boyer_PRL_1980,Nelson_PRB_2017}.
Ionic conduction mechanisms proposed based on such analyses
\cite{Boyer_PRL_1980} are supported by molecular dynamics calculations
\cite{Zhou_SSC_1996}. Soft phonon modes have also been used to
rationalise self-diffusive behaviour, such as that recently discussed
in high-$PT$ iron under Earth-core conditions
\cite{Belonoshko_NGeo}. Experimental neutron-diffraction data
suggestive of soft phonon mode behaviour has been reported in
superionic copper selenide \cite{Danilkin_JPSJ_2010}.

We suggest here that simple descriptors, such as phonon frequencies,
offer a viable means by which to screen candidate materials for
superionic behaviour. The vast majority of structure prediction
studies on new materials proceed by first relaxing candidate
structures using DFT, then using quasiharmonic lattice dynamics to
re-assess the stabilities of low-enthalpy crystal structures, or to
check for dynamic stability
\cite{Nelson_PCCP_2015,Mayo_LiSn,Shamp_JPCL_2015}.
This approach is suitable for high-throughput calculations, and
numerous predictions made using these techniques have been
experimentally verified
\cite{Errea_PRL_2015,Errea_Nature_2016,Li_PRB_2016,Marbella_JACS_2018}. A
wealth of information about harmonic phonons is therefore obtained as
a by-product of structure prediction. This data could be combed for
low-energy or soft modes whose eigenvectors can be identified with
creating disorder, as discussed for $P\overline{6}2m$-CaF$_2$,
$\alpha$-CaF$_2$, CeO$_2$, Li$_2$O and $\beta$-PbF$_2$ in this
work. Frequencies of low-energy optical phonons at $\Gamma$ can also
be examined; Wakamura \cite{Wakamura_PRB_1997} has demonstrated a
strong correlation between these frequencies and the activation
energies required for superionic conduction. Moving away from phonon
frequencies, other descriptors such as dielectric constants and ionic
sizes \cite{Bachman_ChemRev}, elastic constants, or the Lindemann
criterion \cite{Monserrat_PRL_2018} could prove useful in identifying
potential superionics.

Directly screening large numbers of candidate materials using AIMD
calculations is prohibitively expensive. Efforts are ongoing to
substantially reduce the cost of AIMD simulations
\cite{Kahle_PRM_2018}, however, its computational cost remains very
high. Descriptors, such as those discussed here, are suggestive of
superionic behaviour and could be used as a first step to shortlist a
large set of candidate materials or crystal structures for
superionicity, after which molecular dynamics simulations can then be
carried out. In this work, carrying out AIMD simulations with 864-atom
simulation cells on a \textit{single} CaF$_2$ polymorph
($P\overline{6}2m$-CaF$_2$) required about an order of magnitude more
computing time than that used to search the entire Ca$-$F structure
space in our original work \cite{Nelson_PRB_2017}. The latter approach
has the added benefit of identifying new stable stoichiometries, such
as CaF$_3$ \cite{Nelson_PRB_2017}, and BaF$_3$ and BaF$_4$
\cite{Luo_JPCC_2018} (in the case of superionic BaF$_2$) which may
themselves be candidate superionic materials. It is not necessarily
the case that all superionic systems will require large simulation
cells as was the case in this work, but it is difficult to know this
\textit{a priori}.

\section{\label{sec:Conclusions}Conclusions}
$P\overline{6}2m$-CaF$_2$, a polymorph suggested to be stable at high
temperature and pressure \cite{Nelson_PRB_2017}, undergoes a \mbox{type-I}
superionic phase transition to a bcc superionic state,
$Im\overline{3}m$-CaF$_2$. We have observed this transition in
constant-stress \textit{NPT} simulations working at \mbox{2650 K} and
\mbox{20 GPa}. The ionic conductivity is calculated to be in the
neighbourhood of $\sigma\sim 1\:\Omega^{-1}\mbox{cm}^{-1}$ in
$Im\overline{3}m$-CaF$_2$ at 20 GPa and 3000 K.

Modelling the $P\overline{6}2m$-CaF$_2$ phase at high temperature is
difficult. Careful convergence tests need to be carried out to ensure
that appropriately sized simulation cells are used. For
$P\overline{6}2m$-CaF$_2$ at \mbox{2500 K} and \mbox{20 GPa}, the use
of too small a simulation cell leads to the prediction of dominant
calcium, rather than fluorine, diffusion $-$ a result that is both
qualitatively and quantitatively incorrect. Finite-size effects such
as this need to be routinely checked for, and avoided, in molecular
dynamics simulations. Where empirical potentials are available, we
suggest carrying out such convergence tests using classical MD with a
large simulation cell as a benchmark, before any AIMD simulations are
performed. If appropriate force fields are not available, a series of
AIMD tests can still be carried out on small cells to check that
diffusion coefficients are converged.

The softening of phonon modes at the Brillouin zone $X$ point are
investigated for $\alpha$-CaF$_2$, CeO$_2$, $\beta$-PbF$_2$ and
Li$_2$O as a function of volume. Within the uncertainty due to the
treatment of exchange-correlation (or choice of pair potential in the
case of Li$_2$O), these compounds exhibit a harmonic phonon
instability at fractional volumes $V/V_0$ corresponding to a
superionic phase transition. We have discussed the utility of
descriptors, such as soft phonon frequencies, in predicting superionic
behaviour.

\section*{Acknowledgements}
The authors acknowledge useful discussions with Dr S.~Hull, Dr
B.~Monserrat, Dr B.~Karasulu, and Prof.~A.~Walsh. Calculations were
carried out using the \textsc{archer} facility of the U.K.'s national
high-performance computing service, for which access was obtained via
the UKCP consortium (Grant No.~EP/P022596/1).

\pagebreak
\widetext

\large 

\centerline{\textbf{Supplementary material for:}}

\vspace{0.1cm}

\centerline{\textbf{High-pressure CaF$_2$ revisited: a new
    high-temperature phase}}
\centerline{\textbf{and the role of phonons in the search for
    superionic conductivity}}

\vspace{0.1cm}

\centerline{\textbf{J.~R.~Nelson,$^{1,*}$ R.~J.~Needs,$^1$ and C.~J.~Pickard$^{2,3}$}}

\vspace{0.1cm}

\normalsize
\centerline{\textit{$^1$Theory of Condensed Matter Group, Cavendish Laboratory,}}

\centerline{\textit{J.~J.~Thomson Avenue, Cambridge CB3 0HE, United Kingdom}}

\centerline{\textit{$^2$Department of Materials Science and Metallurgy, University of
Cambridge,}}

\centerline{\textit{27 Charles Babbage Road, Cambridge CB3 0FS, United Kingdom}}

\centerline{\textit{$^3$Advanced Institute for Materials Research, Tohoku University, }}

\centerline{\textit{2-1-1 Katahira, Aoba, Sendai, 980-8577, Japan}}

\vspace{0.1cm}

\centerline{$^{*}$Email: \texttt{jn336@cam.ac.uk}}

\subsection*{Effect of cell size and shape on mean-square displacements for $P\overline{6}2m$-CaF$_2$} 
Calculations here use classical molecular dynamics via the
\textsc{lammps} code \cite{LAMMPS_2} alongside the Buckingham pair
potentials discussed in Ref.~\cite{Faraji_PRB_2017_2} for calcium and
fluorine. The \textit{NVT} ensemble is used with \mbox{$T$ = 2500
  K}. $P\overline{6}2m$-CaF$_2$ is set orthorhombically in a cell with
$a=5.9$ \AA, $b= 10.2$ \AA\ and $c= 3.5$ \AA, which yields a pressure
near 20 GPa at \mbox{2500 K}. In the figures below,
$n\times m\times l$ signifies the supercell size referred to this
orthorhombic cell. $D_F$ is the fluorine diffusion coefficient, and
$N$ the number of atoms. Uncertainties in the MSDs for Ca and F,
indicated by light blue and orange shaded regions, are obtained by
averaging across 100 trajectories, with each trajectory starting from
different initial velocities. The exception to this is for $N$=15,552,
for which 20 trajectories instead of 100 are used for averaging.

\begin{figure*}[htp]
\centering
\begin{minipage}{.5\textwidth}
  \centering
  \includegraphics[width=.9\linewidth]{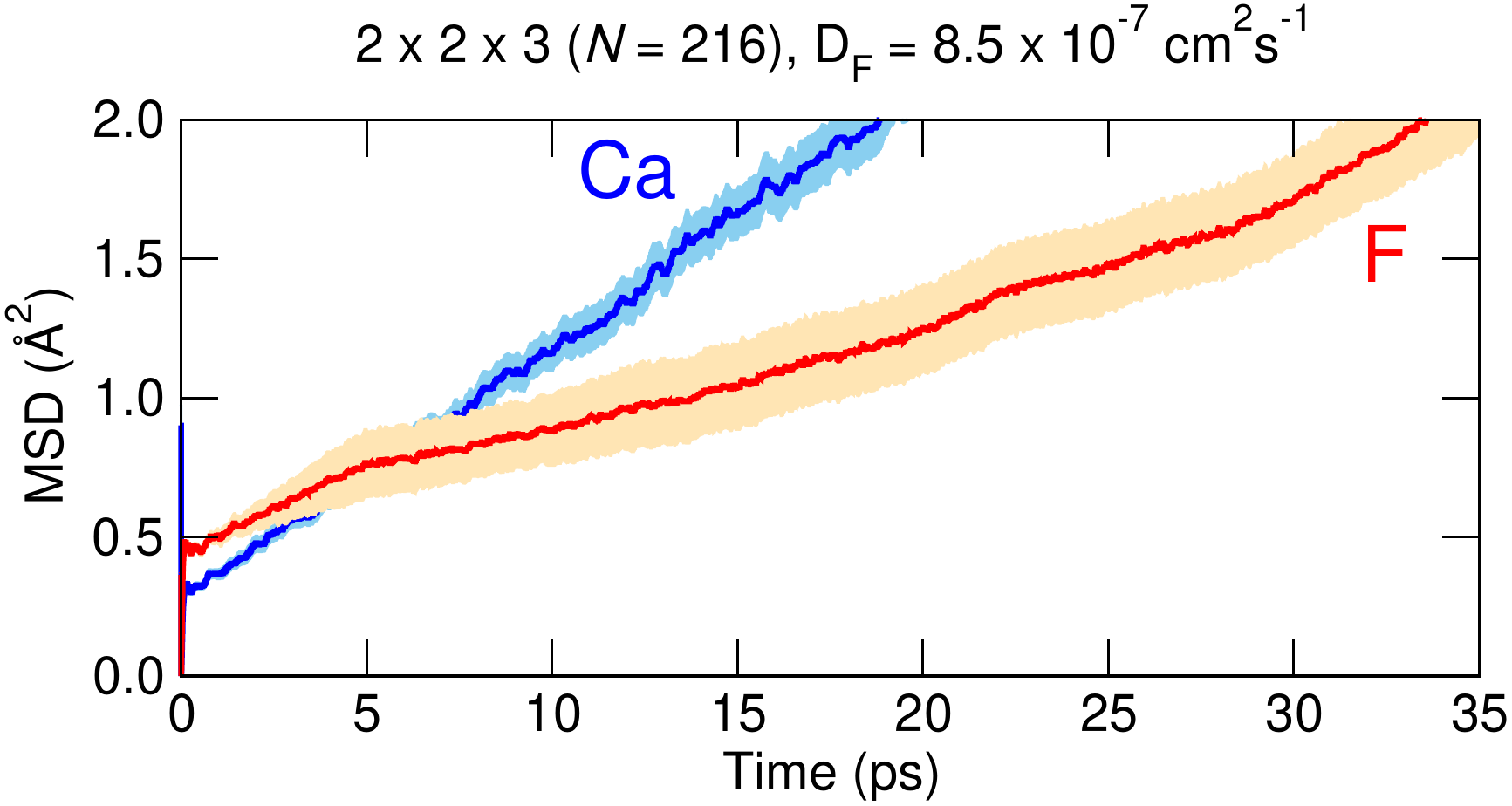} \\ \vspace{0.5cm}
  \includegraphics[width=.9\linewidth]{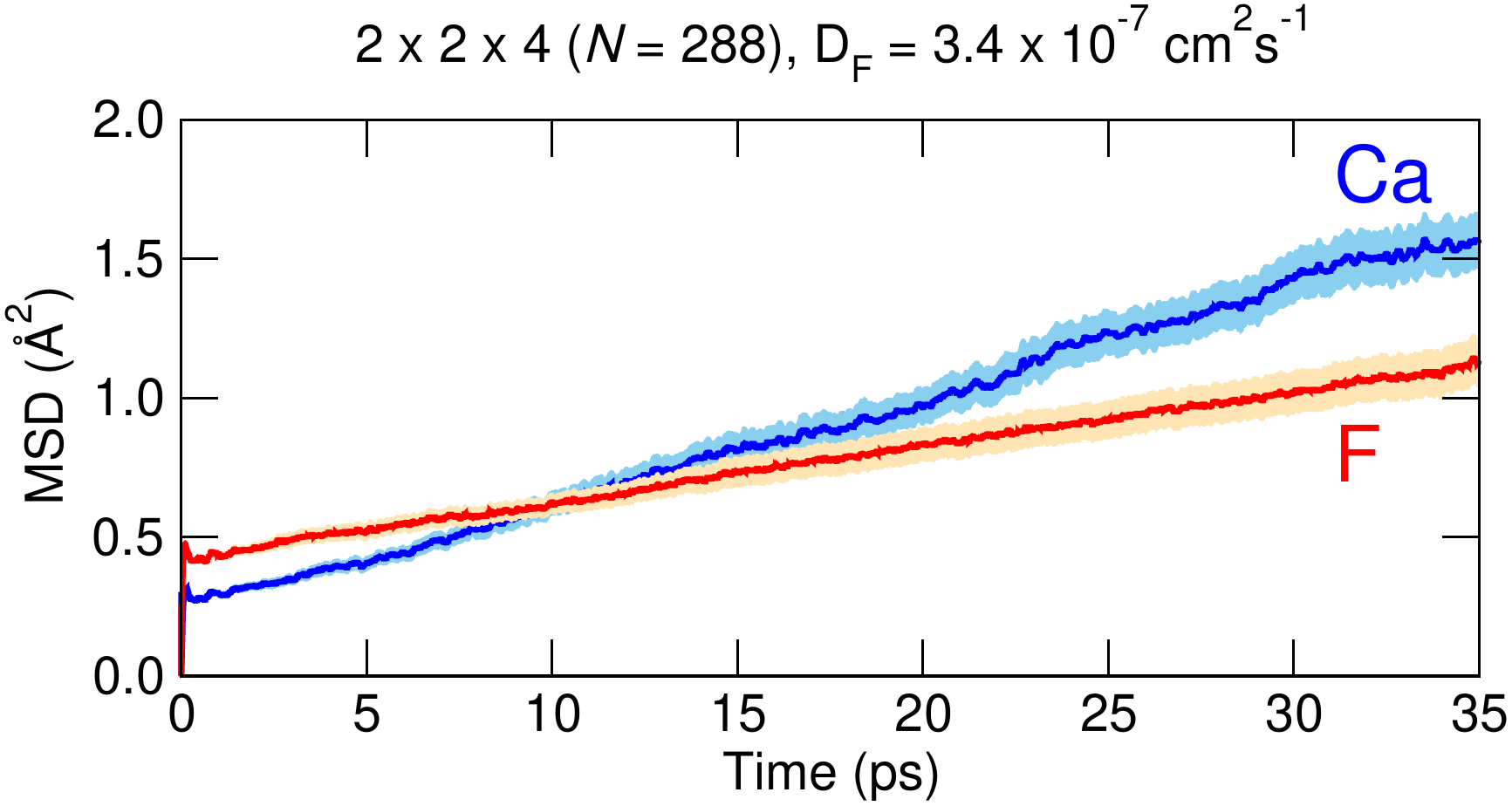} \\ \vspace{0.5cm}
  \includegraphics[width=.9\linewidth]{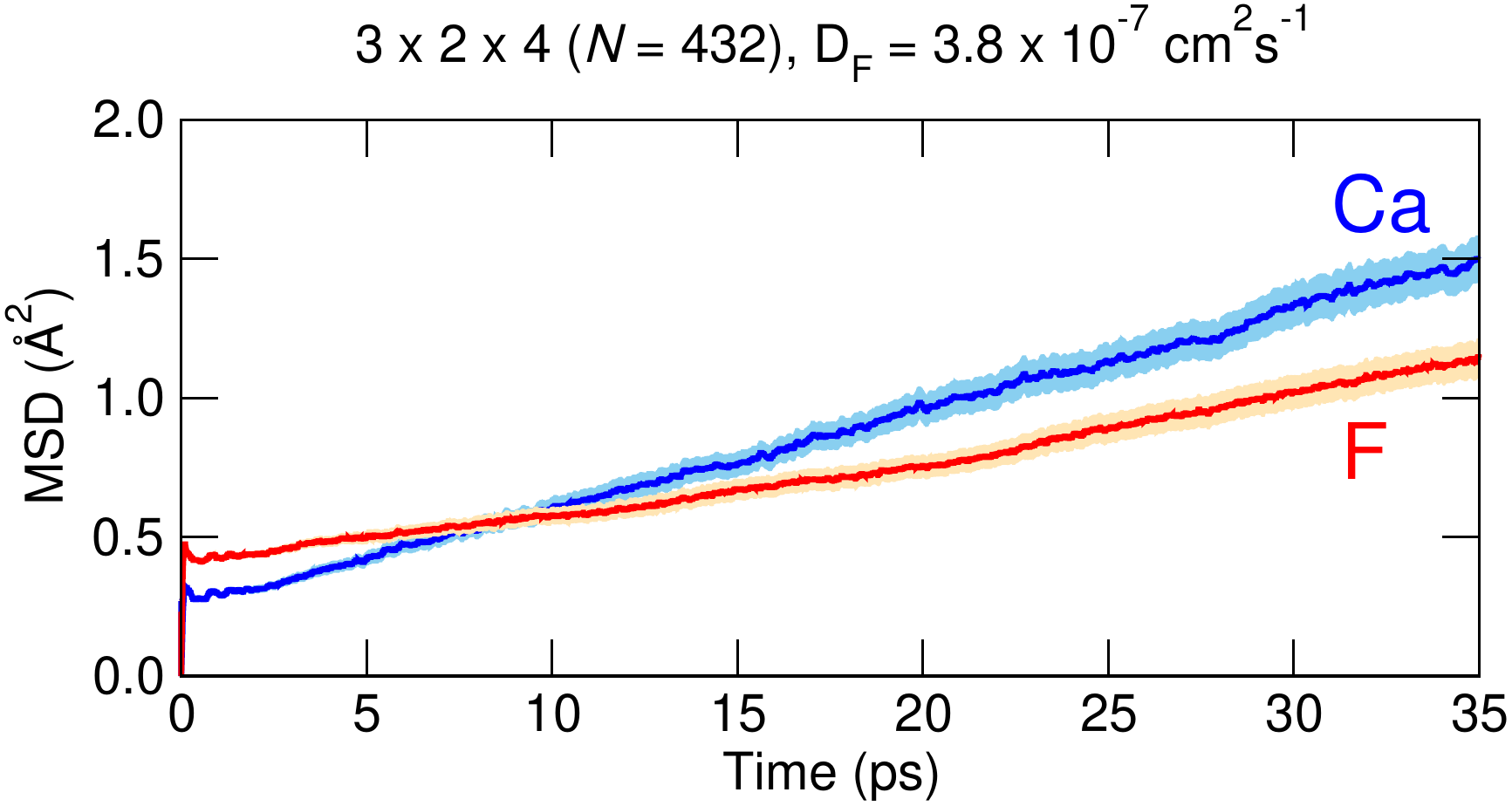} \\ \vspace{0.5cm}
  \includegraphics[width=.9\linewidth]{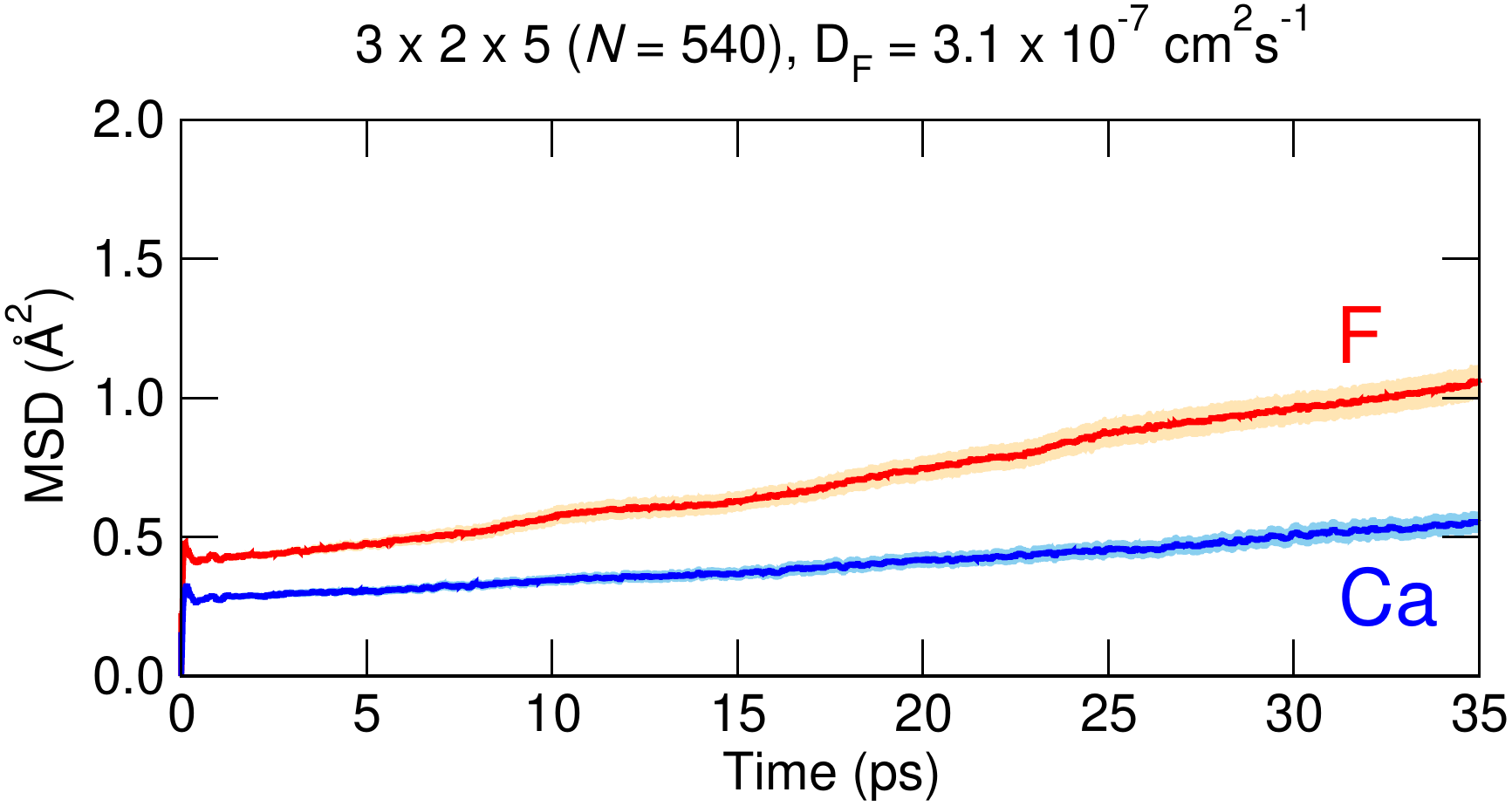} \\ \vspace{0.5cm}
  \includegraphics[width=.9\linewidth]{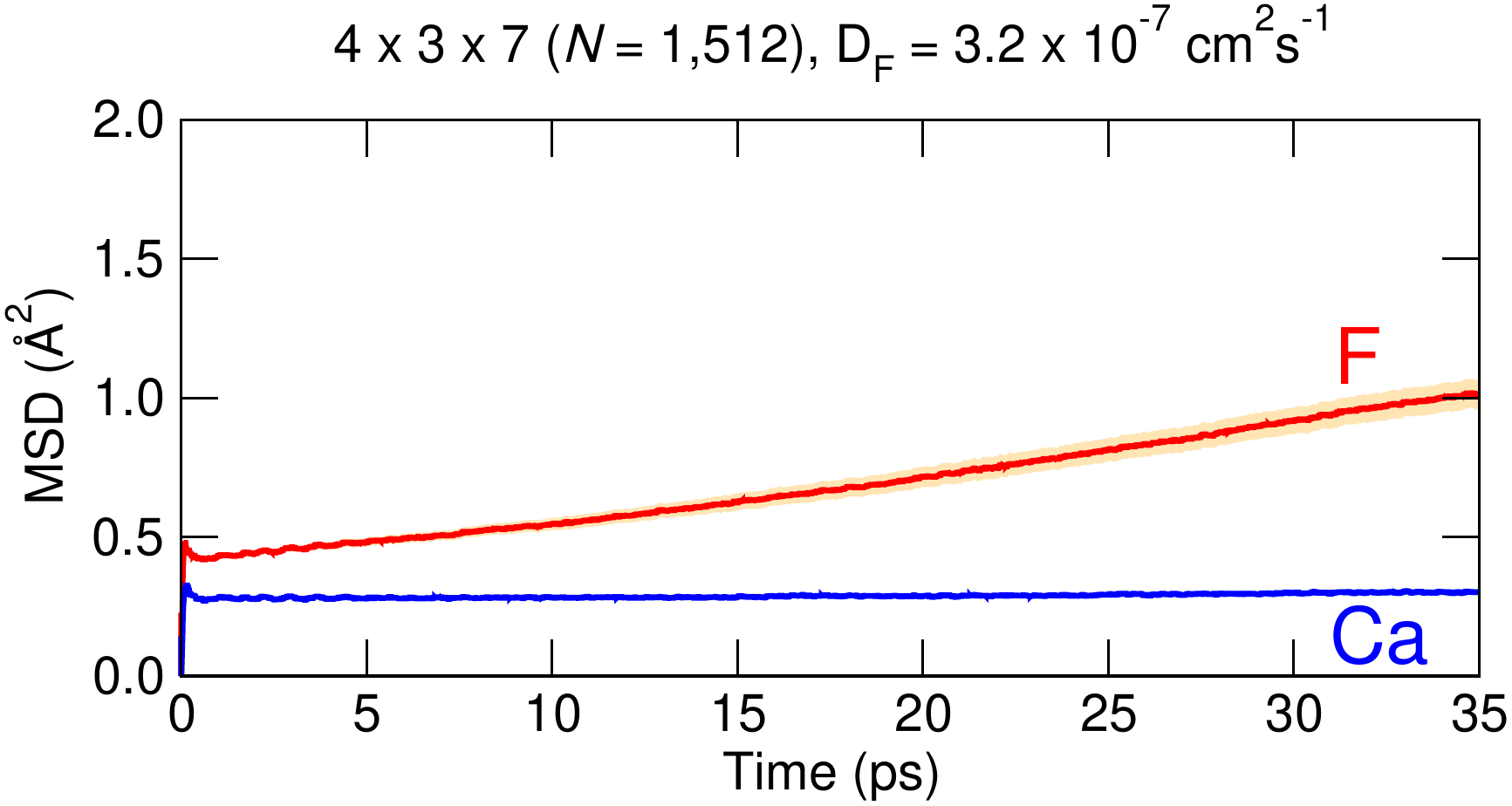}
\end{minipage}%
\begin{minipage}{.5\textwidth}
  \centering
  \includegraphics[width=.9\linewidth]{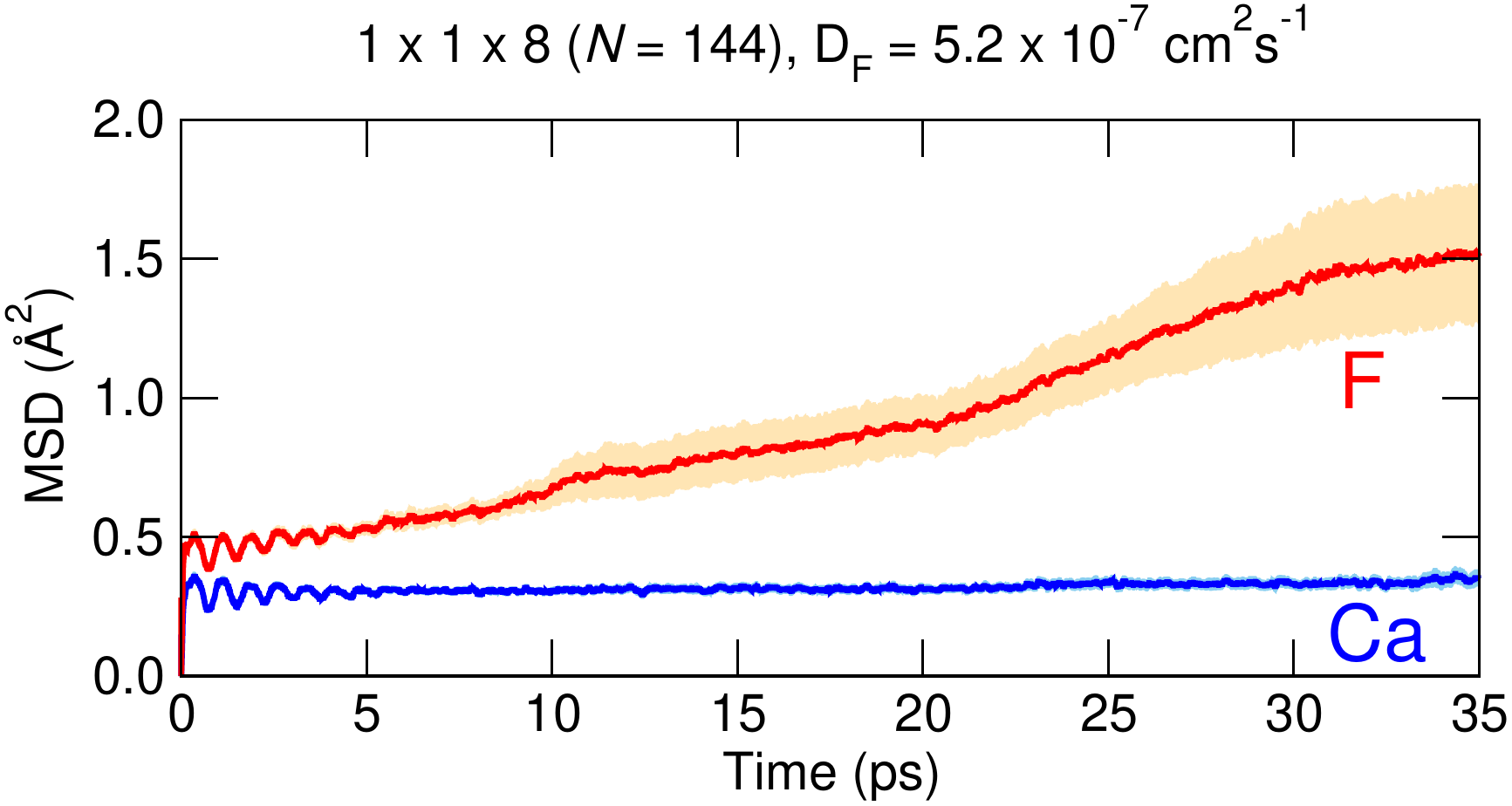} \\ \vspace{0.5cm}
  \includegraphics[width=.9\linewidth]{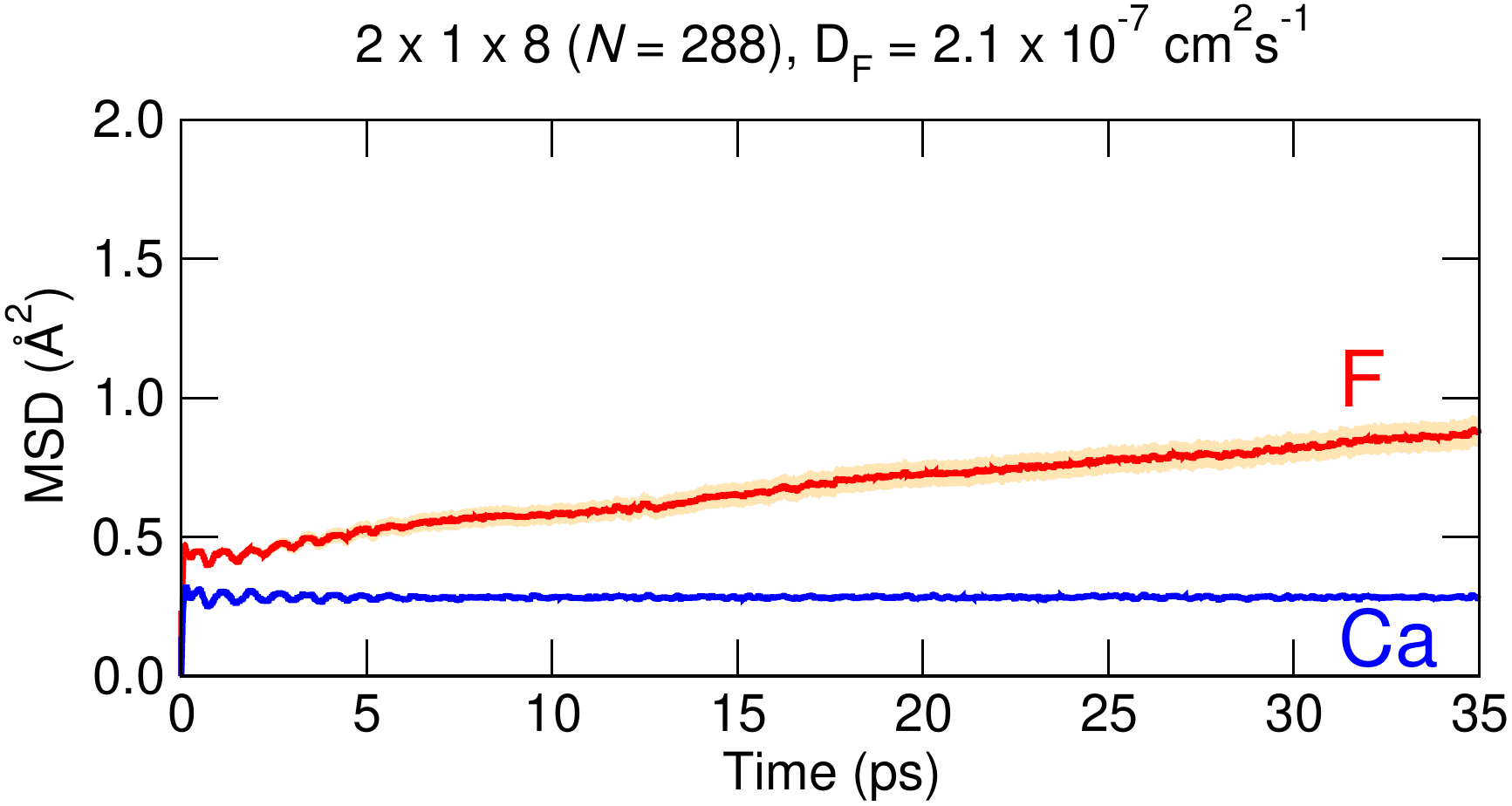} \\ \vspace{0.5cm} 
  \includegraphics[width=.9\linewidth]{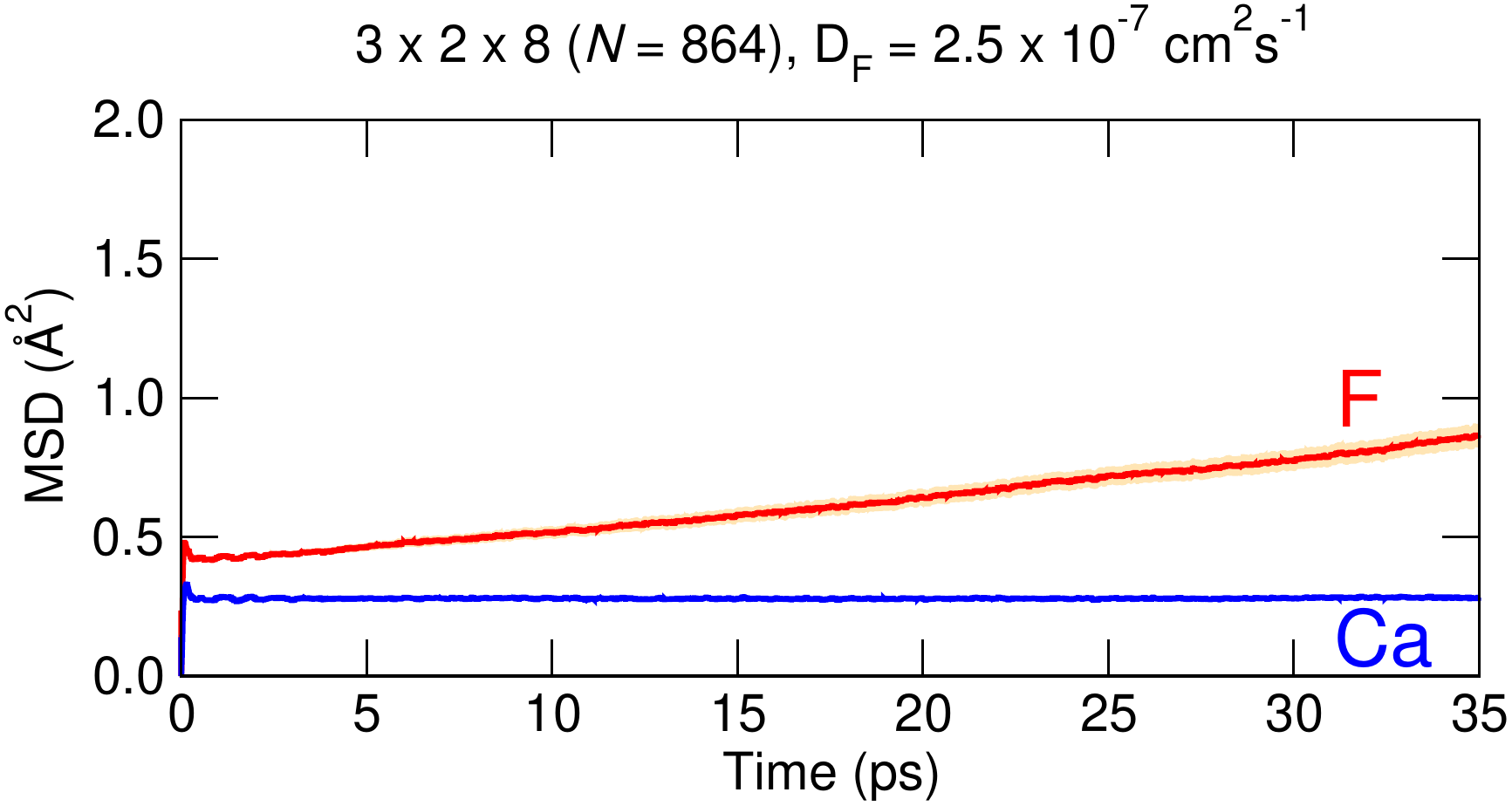} \\ \vspace{0.5cm} 
  \includegraphics[width=.9\linewidth]{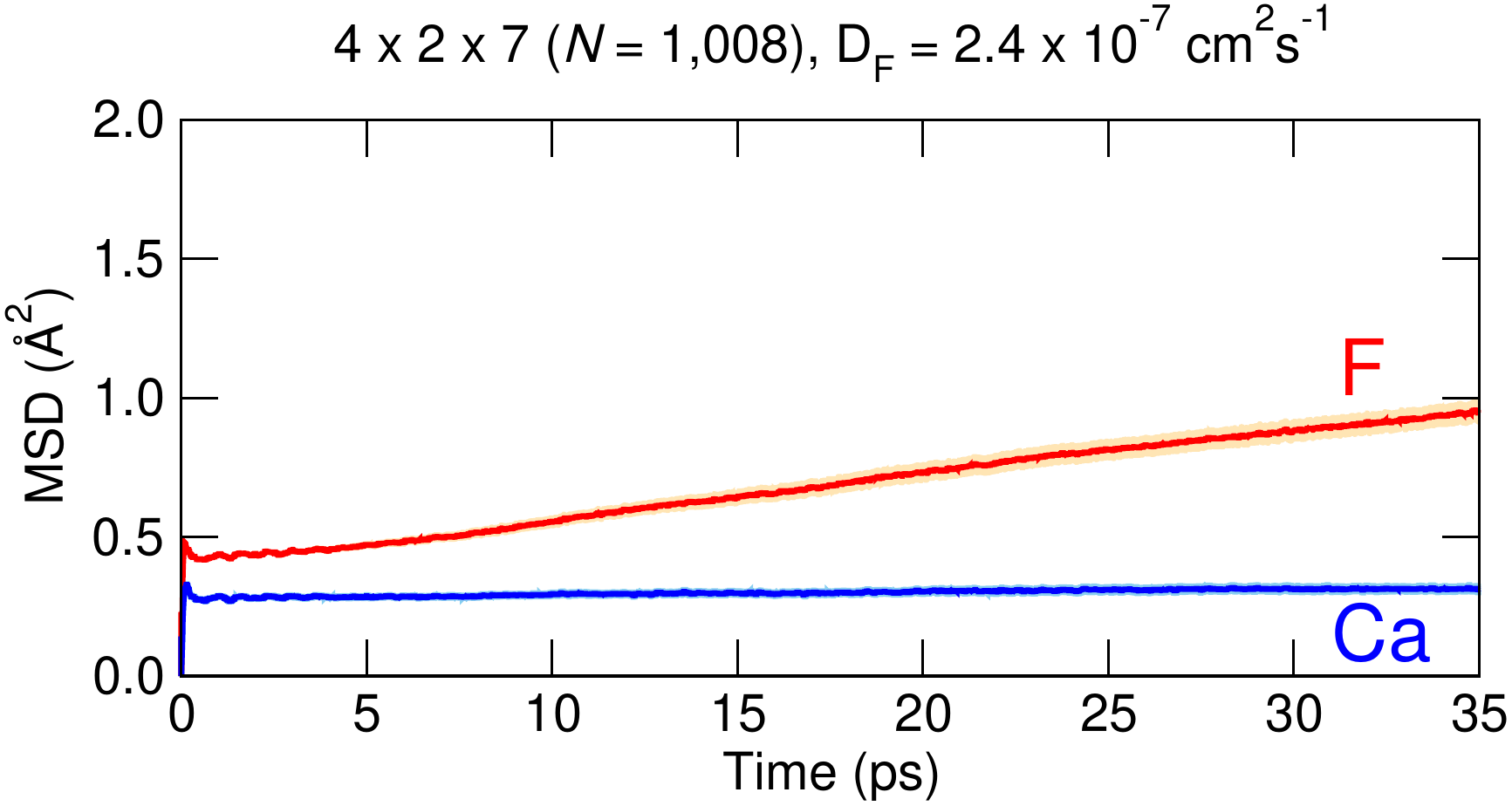} \\ \vspace{0.5cm} 
  \includegraphics[width=.9\linewidth]{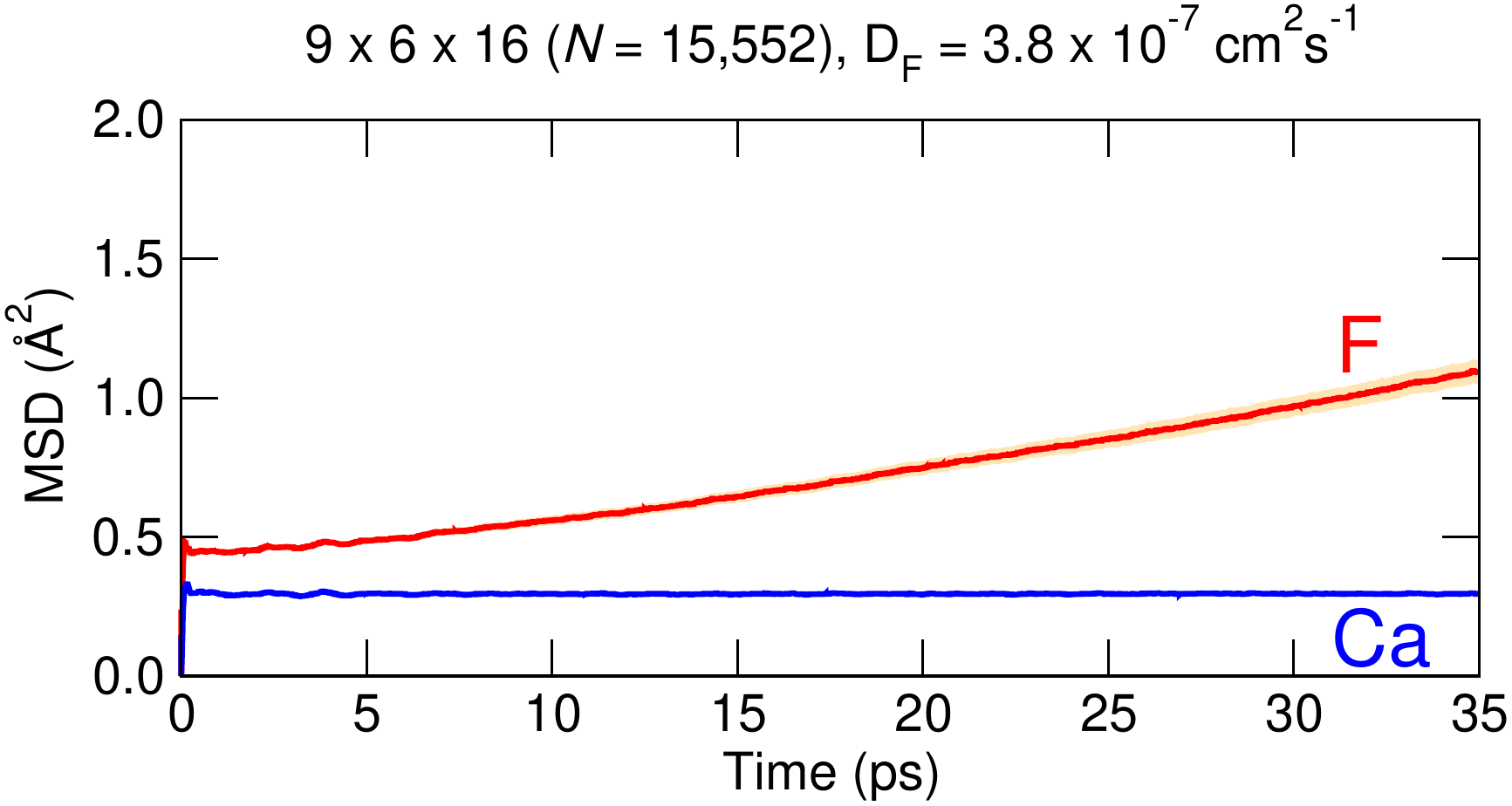} 
\end{minipage}
\caption{\label{fig:fig1}Mean-squared displacements of Ca and F ions
  in $P\overline{6}2m$-CaF$_2$ at \mbox{2500 K} and \mbox{20 GPa}, for
  a variety of different simulation cell shapes and sizes.}
\end{figure*}

\end{document}